\documentclass[12pt]{iopart}

\usepackage{graphicx}
\usepackage{enumerate}
\usepackage{multicol}
\usepackage{color}
\usepackage{amssymb}
\usepackage{pifont}

\begin{document}

\title[1D modelling of long-legged divertor]{Benchmarking of a 1D Scrape-off layer code SOLF1D with SOLPS and its use in modelling long-legged divertors}


\author{E Havl\'i\v ckov\'a$^1$, W Fundamenski$^2$, F Subba$^3$, D Coster$^4$, M Wischmeier$^4$, G Fishpool$^1$}
\address{$^1$ EURATOM/CCFE Fusion Association, Culham Science Centre, Abingdon, Oxon, OX14 3DB, United Kingdom}
\address{$^2$ Imperial College of Science, Technology and Medicine, London, UK}
\address{$^3$ Dipartimento Energia, Politecnico di Torino, Torino, ITALY}
\address{$^4$ Max-Planck Institut f\"ur Plasmaphysik, EURATOM Association, Garching, Germany}

\ead{eva.havlickova@ccfe.ac.uk}

\begin{abstract}
A 1D code modelling SOL transport parallel to the magnetic field (SOLF1D) is benchmarked with 2D simulations of MAST-U SOL performed via the SOLPS code for two different collisionalities. Based on this comparison, SOLF1D is then used to model the effects of divertor leg stretching in 1D, in support of the planned Super-X divertor on MAST. The aim is to separate magnetic flux expansion from volumetric power losses due to recycling neutrals by stretching the divertor leg either vertically ($\nabla B_{\parallel}=0$ in the divertor) or radially ($B\propto 1/R$). 

\end{abstract}

\maketitle

\section{Introduction}


The divertor geometry is one of the important aspects influencing the scrape-off layer (SOL) performance \cite{Loarte}. As conventional divertor concepts might be insufficient to handle power exhaust in future devices, novel magnetic configurations such as Snow-Flake divertor or Super-X divertor (SXD) are considered \cite{Ryutov,Valanju}. In SXD, which is investigated here, a reduction of plasma temperatures in the divertor and energy fluxes to the targets is gained by magnetic flux expansion induced by stretching the divertor leg to larger radius and reducing poloidal magnetic field in the divertor. Secondly, longer connection length $L_{\parallel}$ and closed design of the divertor support power removal caused by plasma-neutral interaction and radiation.  

As part of preparations for the planned SXD on MAST \cite{SXDMast}, numerical investigations of effects associated with long-legged divertor geometry are undertaken using SOL transport codes. In this paper, 1D studies performed with the SOLF1D code are presented, while 2D effects have been simulated using the  SOLPS code coupled with the Monte Carlo code EIRENE \cite{Eirene}, and will be presented in detail in a separate paper. The 1D model enables a separation of the effects of magnetic flux expansion induced by $\nabla_{\parallel}B$ from cooling by plasma-neutral interaction (both functions of the divertor length) by prescribing any parallel/radial dependence of the magnetic field. Such an approach is flexible enough to enable a large number of scans, contrary to robust 2D codes, where each simulation would require a new equilibrium and the preparation of a new grid. However, the 1D approach is limited in terms of the determination of sources/sinks in the divertor due to atomic processes, which are governed by 2D transport of neutral species in the divertor. The 1D code therefore uses an approximation for these sources and compares it with SOLPS simulations perfomed for two configurations with different divertor leg lengths. These two simulations provide a baseline for the scaling of upstream cross-field transport and recycling divertor sources with increasing $L_{\parallel}$. Alternatively, a 1D neutral model in SOLF1D could be used to describe the recycling sources self-consistently, however still excluding 2D processes. 

In order to link results of the 1D and 2D codes, the plasma transport model in SOLF1D is first benchmarked with the SOLPS5.0 model \cite{Schneider}, section \ref{sec_benchmark}. The code comparison is discussed in detail as several discrepancies have been identified. In the following section, 1D effects of the long-legged divertor are discussed based on SOLF1D results. Finally, the first complete documentation for the SOLF1D code is provided in the appendix.

\section{Model description}

\subsection{SOLF1D model for MAST}
\label{sec_solf1d}

SOLF1D is a one-dimensional code solving plasma transport equations along the magnetic field line ($s_{\parallel}$) in the SOL between two targets. Braginskii-like equations in SOLF1D which are defined in \cite{Eva1} have been generalized to take into account the parallel gradient of the magnetic field $\partial B/\partial s_{\parallel}$, while it is assumed that the magnetic field does not change in time $\partial B/\partial t = 0$. This is done in conformity with generalized fluid equations for parallel transport documented in \cite{Zawaideh1}-\cite{Zawaideh2} or \cite{Wojtek1}. Here, a brief description of the model needed for the code benchmark is provided, while more complete documentation of the SOLF1D model can be found in the appendix. 

The set of equations solved in the code includes the continuity and momentum equations for plasma density $n$ and parallel ion velocity $u_{\parallel}$, and energy equations for electron and ion temperatures $T_{\rm e}$ and $T_{\rm i}$ 
\begin{eqnarray}
\hspace{-2cm}\frac{\partial n}{\partial t}+B\frac{\partial}{\partial s_{\parallel}}\left(\frac{nu_{\parallel}}{B}\right)=S^n, \label{eq_continuity} \\
\hspace{-2cm}\frac{\partial}{\partial t}(m_{\rm i}nu_{\parallel})+B\frac{\partial}{\partial s_{\parallel}}\left(\frac{m_{\rm i} n u_{\parallel}^2}{B}\right)+B^{\frac{3}{2}}\frac{\partial}{\partial s_{\parallel}}\left(B^{-\frac{3}{2}} \delta p_{\rm i}\right)=-\frac{\partial p_{\rm i}}{\partial s_{\parallel}}+enE_{\parallel}+R_{\rm \parallel,i}+m_{\rm i}S_{\parallel}^u, \label{eq_momentum} \\
\hspace{-2cm}\frac{\partial}{\partial t}\left(\frac{3}{2}nkT_{\rm e}\right)+B\frac{\partial}{\partial s_{\parallel}}\left[\frac{1}{B}\left(\frac{5}{2}nkT_{\rm e}u_{\parallel}+q_{\rm \parallel,e}\right)\right]=-enu_{\parallel}E_{\parallel}+u_{\parallel}R_{\rm \parallel,e}+Q_{\rm e}+S_{\rm e}^E, \label{eq_energy1} \\
\hspace{-2cm}\frac{\partial}{\partial t}\left(\frac{3}{2}nkT_{\rm i}+\frac{1}{2}m_{\rm i}nu_{\parallel}^2\right)+B\frac{\partial}{\partial s_{\parallel}}\left[\frac{1}{B}\left(\frac{5}{2}nkT_{\rm i}u_{\parallel}+\frac{1}{2}m_{\rm i}nu_{\parallel}^3+q_{\rm \parallel,i}+u_{\parallel}\delta p_{\rm i} \right)\right]= \label{eq_energy2} \\
enu_{\parallel}E_{\parallel}+u_{\parallel}R_{\rm \parallel,i}+Q_{\rm i}+S_{\rm i}^E.  \nonumber
\end{eqnarray}
We assume that the electron density is equal to the ion one $n_{\rm e}=n_{\rm i}=n$. Further, zero net parallel current is assumed $j_{\parallel}=0$ ($u_{\rm \parallel,e}=u_{\rm \parallel,i}$) together with the generalized Ohm's law for the electron momentum $enE_{\parallel}=-\partial p_{\rm e}/\partial s_{\parallel}+R_{\rm \parallel,e}$, where $E_{\parallel}$ is the parallel electric field and $R_{\rm e}$ $(R_{\rm e}=-R_{\rm i})$ are the thermal and friction forces. $S^n$, $S^u$, $S_{\rm e}^E$ and $S_{\rm i}^E$ are sources due to collisions with neutrals and cross-field transport sources. $Q_{\rm e}$ and $Q_{\rm i}$ is the heating due to electron-ion collisions, $q_{\rm \parallel,e}=-\kappa_e \partial (kT_{\rm e})/\partial s_{\parallel}$ and $q_{\rm \parallel,i}=-\kappa_i \partial (kT_{\rm i})/\partial s_{\parallel}$ are the electron and ion heat fluxes, $p_{\rm e}$ and $p_{\rm i}$ is the electron and ion static pressure and $\delta p_{\rm i}=-\eta_{\parallel}B^{-1/2}\partial/\partial s_{\parallel} (B^{1/2}u_{\parallel})$ reflects the anisotropic part of the pressure tensor for ions (electrons are assumed Maxwellian $\delta p_{\rm e}=0$). 

Boundary conditions at the targets are based on the sheath theory and include the Bohm criterion for the ion speed $u_{\parallel}=c_{\rm s}$ and energy fluxes controlled by sheath heat transmission factors $Q_{\rm \parallel,e}=\delta_{\rm e}nkT_{\rm e}c_{\rm s}$ and $Q_{\rm \parallel,i}=\delta_{\rm i}nkT_{\rm i}c_{\rm s}$. $Q_{\rm \parallel}$ denotes the total energy flux, $Q_{\rm \parallel,e}=5/2nkT_{\rm e}u_{\parallel}+q_{\rm \parallel,e}$ and $Q_{\rm \parallel,i}=5/2nkT_{\rm i}u_{\parallel}+1/2m_{\rm i}nu_{\parallel}^3+q_{\rm \parallel,i}+u_{\parallel}\delta p_{\rm i}$. The sound speed is defined as $c_{\rm s}=\sqrt{k(T_{\rm e}+T_{\rm i})/m_{\rm i}}$. The sheath heat transmission coefficients are fixed to constant values of $\delta_{\rm e}=5.0$ and $\delta_{\rm i}=3.5$. The density at the target is obtained by an extrapolation.

\subsection{SOLPS equations and their reduction to 1D}
\label{sec_solps}

The SOLPS model is based on the Braginskii equations \cite{Schneider} solved in the poloidal geometry, i.e. assuming toroidal symmetry. The conservation equations are written in curvilinear coordinates coinciding with the magnetic topology. The equations solved in SOLPS are  documented in \cite{Schneider}. Here, SOLPS5.0 without drifts activated is used. We also assume that $j_{\parallel}=0$ and the potential equation is not followed. Parallel transport is described by the Braginskii model with the use of viscous and heat flux limiters. Note that the change from the default Balescu to Braginskii transport coefficients and setting $j_{\parallel}=0$ has no visible effect on the solution for the MAST-U cases presented in section \ref{sec_benchmark}, while it assures a consistency with the 1D model. In addition, there is no impurity present in the simulation. The plasma temperature at the outer target is increased by 30\% in comparison with a simulation including C sputtering and assuming constant chemical sputtering rate of 1\% and physical sputtering yield calculated from Roth-Bogdansky formula (this simulation is not shown here). SOLPS equations in \cite{Schneider} can be simplified into the following form
\begin{equation}
\hspace{-2cm}\frac{\partial n}{\partial t}+\frac{1}{\sqrt{g}}\frac{\partial}{\partial x}\left(\frac{\sqrt{g}}{h_{x}}n u_x\right)+\frac{1}{\sqrt{g}}\frac{\partial}{\partial y}\left(\frac{\sqrt{g}}{h_{y}}n u_y\right)=S^{(n)} \label{eq_solps1} 
\end{equation}
with $u_x=b_xu_{\parallel}+b_zu_{\perp}$ and anomalous radial transport reflected in the radial velocity as $u_y=-(D^n/nh_y) \partial n/\partial y$, $b_zu_{\perp}=-(D^n/nh_x) \partial n/\partial x$,
\begin{eqnarray}
\hspace{-2cm}m_{\rm i}\left[\frac{\partial nu_{\parallel}}{\partial t}+\frac{1}{h_z\sqrt{g}}    \frac{\partial}{\partial x}\left(\frac{h_z\sqrt{g}}{h_x}nu_xu_{\parallel}\right)+\frac{1}{h_z\sqrt{g}}\frac{\partial}{\partial y}\left(\frac{h_z\sqrt{g}}{h_y}nu_yu_{\parallel}\right)\right] \\
-\frac{4}{3}b_xB^{\frac{3}{2}}\frac{\partial}{h_x\partial x}\left(\frac{\eta_0b_x}{B^2}\frac{\partial B^{\frac{1}{2}}u_{\parallel}}{h_x\partial x}\right)-B^{\frac{3}{2}}b_x\frac{\partial}{h_x \partial x}\left(\frac{b_x}{\nu_{\rm ii}B^2}\frac{\partial B^{\frac{1}{2}}q_{\rm \parallel,i}}{h_x\partial x}\right)= \nonumber \\ 
-\frac{b_x}{h_x}\frac{\partial p_{\rm i}}{\partial x}-b_x\frac{en}{h_x}\frac{\partial \phi}{\partial x}+R_{\rm \parallel,i}+S^{(u)}_{\parallel} \label{eq_solps2} \nonumber 
\end{eqnarray}
with an additional parallel viscosity driven by the ion heat flux $q_{\rm \parallel,i}=-\kappa_{\rm \parallel,i}b_x\partial T_{\rm i}/h_x\partial x$ which is not included in the standard Braginskii model,
\begin{equation}
\hspace{-2cm} \frac{3}{2}\frac{\partial nkT_{\rm e}}{\partial t}+\frac{1}{\sqrt{g}}\frac{\partial}{\partial x}\left(\frac{\sqrt{g}}{h_{x}}q_{x,{\rm e}}\right)+\frac{1}{\sqrt{g}}\frac{\partial}{\partial y}\left(\frac{\sqrt{g}}{h_{y}}q_{y,{\rm e}}\right)+\frac{nkT_{\rm e}}{\sqrt{g}}\frac{\partial}{\partial x}\left(\frac{\sqrt{g}}{h_x}b_xu_{\parallel}\right)=Q_{\rm e}+S_{\rm e}^{(E)}, \label{eq_solps3} 
\end{equation}
\begin{eqnarray}
q_{x,{\rm e}}=\frac{3}{2}nkT_{\rm e}b_xu_{\parallel}-\kappa_{\rm \parallel,e}\frac{b_x^2}{h_x}\frac{\partial kT_{\rm e}}{\partial x}-\frac{5}{2}nkT_{\rm e}b_zD_{\rm an}\frac{1}{h_xn}\frac{\partial n}{\partial x}, \nonumber \\ 
q_{y,{\rm e}}=-\frac{5}{2}nkT_{\rm e}D_{\rm an}\frac{1}{h_yn}\frac{\partial n}{\partial y},  \nonumber  
\end{eqnarray}
\begin{equation}
\hspace{-2cm} \frac{3}{2}\frac{\partial nkT_{\rm i}}{\partial t}+\frac{1}{\sqrt{g}}\frac{\partial}{\partial x}\left(\frac{\sqrt{g}}{h_{x}}q_{x,{\rm i}}\right)+\frac{1}{\sqrt{g}}\frac{\partial}{\partial y}\left(\frac{\sqrt{g}}{h_{y}}q_{y,{\rm i}}\right)+\frac{nkT_{\rm i}}{\sqrt{g}}\frac{\partial}{\partial x}\left(\frac{\sqrt{g}}{h_x}b_xu_{\parallel}\right)=Q_{\rm vis}+Q_{\rm i}+S_{\rm i}^{(E)}, \label{eq_solps4} 
\end{equation}
\begin{eqnarray}
q_{x,{\rm i}}=\frac{3}{2}nkT_{\rm i}b_xu_{\parallel}-\kappa_{\rm \parallel,i}\frac{b_x^2}{h_x}\frac{\partial kT_{\rm i}}{\partial x}-\frac{5}{2}nkT_{\rm i}b_zD_{\rm an}\frac{1}{h_xn}\frac{\partial n}{\partial x}, \nonumber \\      
q_{y,{\rm i}}=-\frac{5}{2}nkT_{\rm i}D_{\rm an}\frac{1}{h_yn}\frac{\partial n}{\partial y}     \nonumber, \\
Q_{\rm vis}=\frac{\eta_0}{3}\left(\frac{2b_x}{\sqrt{B}}\frac{\partial u_{\parallel}\sqrt{B}}{h_x\partial x}\right)^2. \nonumber
\end{eqnarray}

It can be shown that Eqs. (\ref{eq_solps1})-(\ref{eq_solps4}) reduce into 1D equations similar to Eqs. (\ref{eq_continuity})-(\ref{eq_energy2}), see below. In the code, the equations are discretized using the finite volume method. For example, the finite volume form of the continuity equation (\ref{eq_solps1}) assuming steady state is 
\begin{equation}
\Delta F^n_x + \Delta F^n_y = S^{(n)}V \label{eq_fvm}
\end{equation}
where on the left side, we sum the particle fluxes $F^n_x$ and $F^n_y$ entering and leaving the cell across the cell boundaries, and the right side represents the total particle net source in s$^{-1}$ on the cell with volume $V$. 

The poloidal flux $F^n_x$ across the cell face is calculated as $F^n_x=nu_{\parallel}AB_x/B$, where $A=2\pi R h$ is the radial area of the cell perpendicular to the poloidal direction with the radial size of the cell $h$ across the flux tube. Because SOLPS does not use a staggered grid for all flow variables, the Rhie-Chow interpolation method \cite{Rhie} is employed to take the cell-centre values of the parallel velocity $u_{\parallel}$ on the cell faces where fluxes are calculated. The cell volume is defined as $V=2\pi R h\Delta x$. Following from Eq. (\ref{eq_fvm}), the poloidal part of the flux divergence can be translated as
\begin{equation}
\frac{\Delta F^n_x}{V}=\frac{1}{A} \frac{\Delta \left(nu_{\parallel}A\frac{B_x}{B}\right)}{\Delta x}=\frac{1}{Rh}\frac{\Delta\left(nu_{\parallel}Rh\frac{B_x}{B}\right)}{\Delta x}.
\end{equation}
Using (i) $B_z\propto 1/R$ and (ii) $hB_x/B_z={\rm const}$ along the flux tube, and replacing $\Delta x$ by $(B_x/B)\Delta s_{\parallel}$, this further yields
\begin{equation}
\frac{\Delta F^n_x}{V}=\frac{1}{Rh}\frac{\Delta\left(nu_{\parallel}Rh\frac{B_x}{B}\right)}{\Delta x}=B_x \frac{\Delta \left(\frac{nu_{\parallel}}{B}\right)}{\Delta x}=B \frac{\Delta \left(\frac{nu_{\parallel}}{B}\right)}{\Delta s_{\parallel}},
\end{equation}
equivalent to the divergence term of a 1D equation 
\begin{equation}
B\frac{\Delta \left(\frac{nu_{\parallel}}{B}\right)}{\Delta s_{\parallel}}=S^n \label{eq_fdm}
\end{equation}
consistent with the one solved in SOLF1D, Eq. (\ref{eq_continuity}). 
The radial part of the flux divergence $\Delta F^n_y/V_{\rm cell}$  from Eq. (\ref{eq_fvm}) appears as a source term in the equation parallel to the magnetic field (\ref{eq_fdm}), $S^n=S^{(n)}-\Delta F^n_y/V$.

Boundary conditions in SOLPS for quantities at the target are not exactly identical to those in SOLF1D. In SOLPS, the density is calculated assuming zero gradient at the target (a comparison with SOLF1D shown later in Fig. \ref{fig_other1}). The parallel ion velocity is set to the sound speed as in SOLF1D, $u_{\parallel}=c_{\rm s}$, using the same definition for $c_{\rm s}$. The target energy fluxes are prescribed as $Q_{\rm \parallel,e}^{in}=3/2nkT_{\rm e}u_{\parallel}+q_{\rm \parallel,e}=\gamma_{\rm e}nkT_{\rm e}c_{\rm s}$ and $Q_{\rm \parallel,i}^{in}=3/2nkT_{\rm i}u_{\parallel}+q_{\rm \parallel,i}=\gamma_{\rm i}nkT_{\rm i}c_{\rm s}$ with $\gamma_{\rm e}=4.0$ and $\gamma_{\rm i}=2.5$ as SOLPS solves the internal energy equation instead of the typical conservative form of the energy balance (hence the index $in$), i.e. the ion kinetic and viscous parts of the energy flux are not included in the boundary condition (see the difference between the codes in Fig. \ref{fig_other1}). We used different notation for the sheath heat transmission coefficients than in section \ref{sec_solf1d} to take account of  different values of the coefficients in SOLPS versus SOLF1D due to different definitions of the sheath energy fluxes. 

\section{Benchmark of codes}
\label{sec_benchmark}

\begin{figure}[!h]
\centerline{\scalebox{0.5}{\includegraphics[clip]{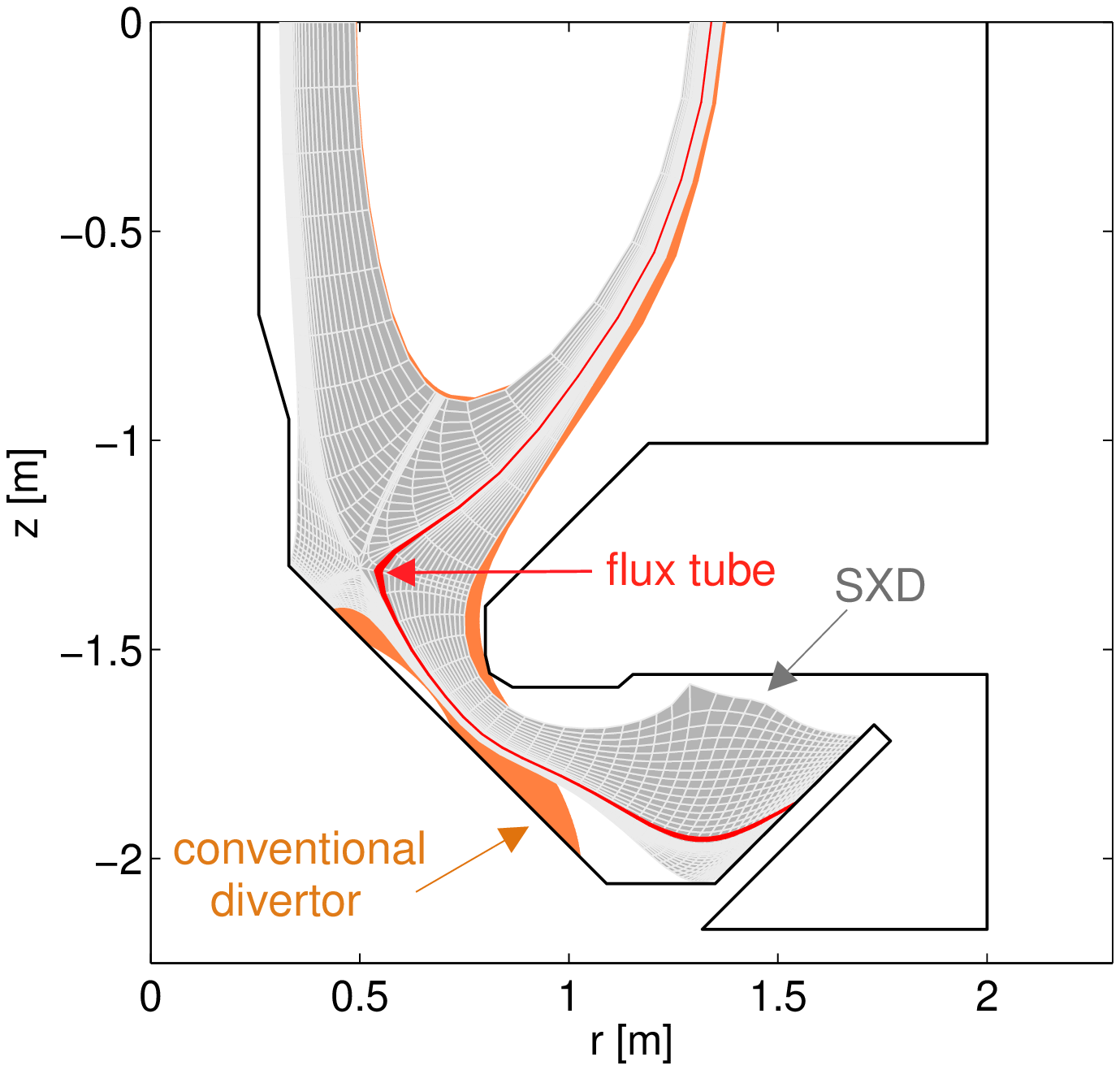}}
\scalebox{0.7}{\includegraphics[clip]{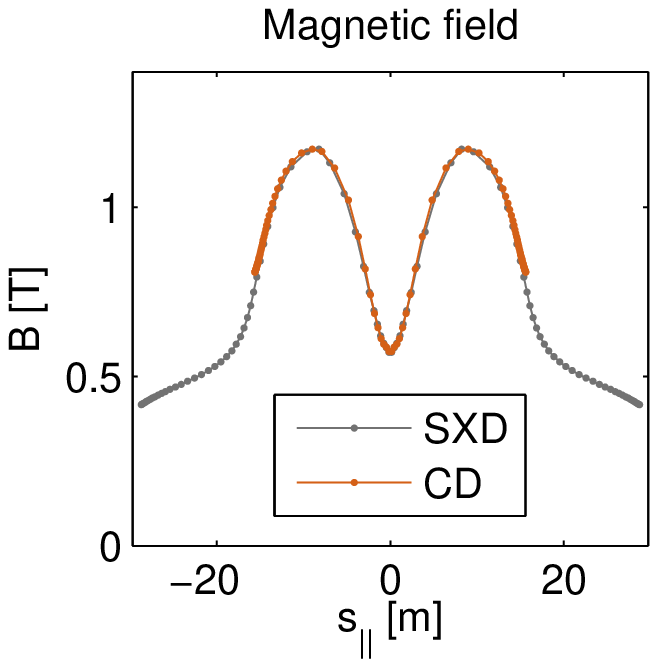}}
}
\caption{On the left, two divertor geometries used for benchmarking the codes -- conventional divertor and SXD. The grid is top/bottom symmetric and the flux tube selected for benchmarking, shown in red, is located at $\Delta r_{\rm sep}\approx 0.5$ mm at the outboard side between top and bottom targets. On the right, the magnetic field along the flux tube is shown for the two divertor configurations.} \label{fig_scheme}
\end{figure}

For benchmarking the codes, two converged SOLPS solutions were selected. Both are for MAST-U, H-mode plasmas, in connected double null magnetic configuration, the first one for the Super-X divertor (SXD) geometry, the second one for the conventional divertor (CD) geometry (Fig. \ref{fig_scheme} left). These cases have been studied in \cite{Eva4}. The input power is $P_{\rm inp}=1.7MW$ (the power crossing the core boundary) and the density at the core boundary is $n_{\rm core}=2.8 \times 10^{19}$ m$^{-3}$. 
Transport coefficients are $D_{\perp}=\chi_{\perp}=1$ m$^2$s$^{-1}$ in the SOL, $D_{\perp}=\chi_{\perp}=2$ m$^2$s$^{-1}$ in the core, but reduced to $D_{\perp}=0.2$ m$^2$s$^{-1}$ and $\chi_{\perp}=0.5$ m$^2$s$^{-1}$ in the pedestal (a region extending 2 cm inside and 0.5 cm outside the separatrix) to enforce a transport barrier.
A flux tube used for the comparison is close to the separatrix (the radial distance from the separatrix is approximately $\Delta r_{\rm sep}\approx 0.5$ mm at the outboard midplane) and it is the flux tube with maximum energy flux at the target. On the right side of Fig. \ref{fig_scheme}, the magnetic field along this flux tube between the top and bottom targets is shown, with a larger drop in SXD as expected from the extension of the divertor to larger radius. 

\begin{figure}[!h]
\centerline{\scalebox{0.7}{\includegraphics[clip]{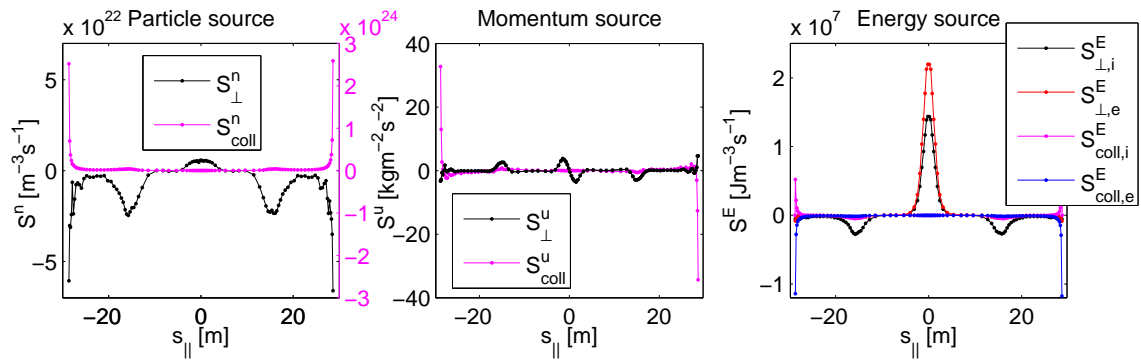}}}
\caption{SOLPS modelling of SXD geometry -- sources of particles, momentum and energy along a flux tube located close to the separatrix between the top and bottom outer targets. Two types of sources are shown -- sources due to transport radially across the flux tube $S_{\perp}$ and sources due to recycling calculated in EIRENE $S_{\rm soll}$.} \label{fig_sources1}
\end{figure}

\begin{figure}[!h]
\centerline{\scalebox{0.7}{\includegraphics[clip]{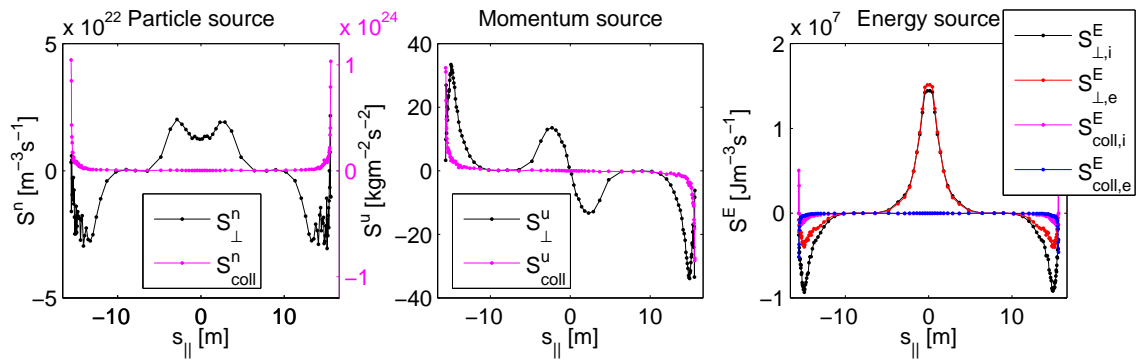}}}
\caption{SOLPS modelling of CD geometry -- sources of particles, momentum and energy along a flux tube.} \label{fig_sources2}
\end{figure}

Sources from EIRENE due to plasma-neutral collisions and sources from SOLPS due to radial transport are used as an input for SOLF1D, together with the magnetic field variation along the flux tube. The sources are displayed in Figs. \ref{fig_sources1}-\ref{fig_sources2}. The energy balance on the flux tube is dominated by the radial transport in the upstream SOL and below the X points. The contribution from the electron cooling due to plasma-neutral interaction close to the targets is small compared to the radial transport, but it is stronger in the SXD case than in the CD case. The particle sources are on the contrary dominated by the ionization of recycled neutrals in front of the targets, and this recycling source is again stronger in the SXD case (larger divertor volume and the connection length, closed divertor, more collisions, smaller temperatures). The radial particle sources are comparable in magnitude and show the same pattern -- a source at the stagnation point and a sink below the X points due to transport from the SOL to the private flux region. 

\subsection{Standard model}

\begin{figure}[!h]
\centerline{\scalebox{0.7}{\includegraphics[clip]{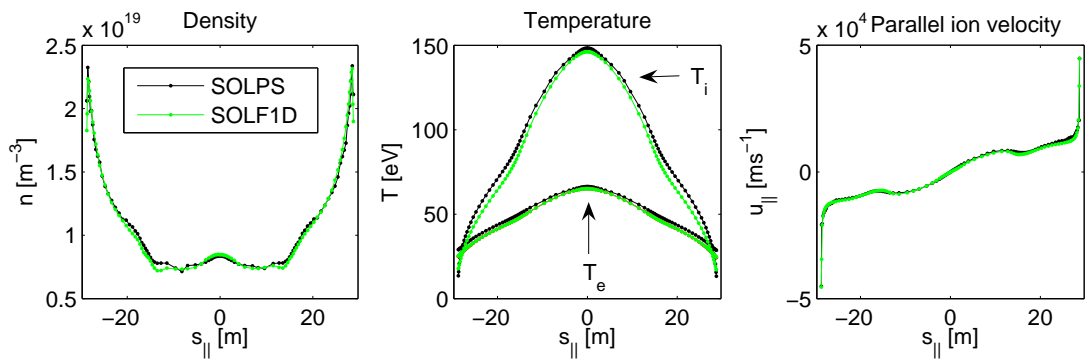}}}
\caption{A flux tube in the SXD grid from Fig. \ref{fig_scheme} -- comparison of SOLF1D and SOLPS solutions.} \label{fig_bench1}
\end{figure}

\begin{figure}[!h]
\centerline{\scalebox{0.7}{\includegraphics[clip]{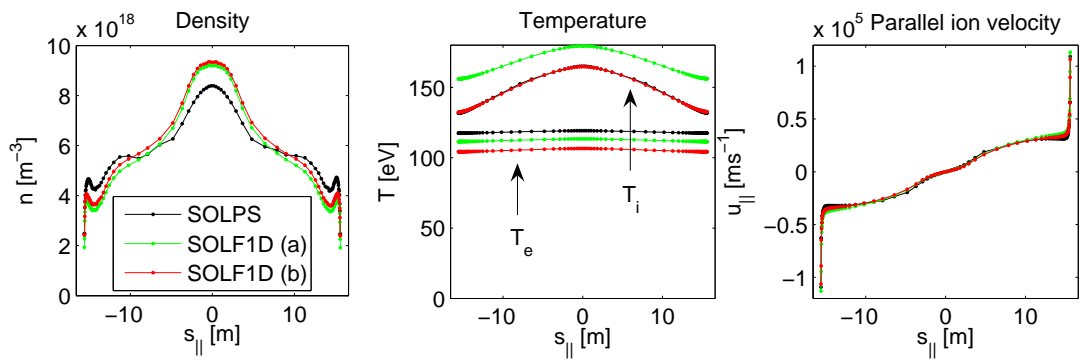}}}
\caption{A flux tube in the CD grid from \ref{fig_scheme} -- comparison of SOLF1D and SOLPS solutions. (a) The standard SOLF1D model as described in section \ref{sec_solf1d}, (b) the SOLF1D model using the SOLPS definition of the boundary condition for $T_{\rm i}$.} \label{fig_bench2}
\end{figure}

Solutions of SOLF1D and SOLPS on a flux tube from Fig. \ref{fig_scheme}, which is in the SXD geometry, and using the sources from Fig. \ref{fig_sources1}, are compared in Fig. \ref{fig_bench1}, showing a very good agreement between the codes. A similar level of agreement has been achieved also for the CD geometry, apart from flux tubes located very close to the separatrix at $\Delta r_{\rm sep}\lessapprox 1$ mm. 
These flux tubes are the subject of further attention in order to identify the origin of the mismatch. An example of the least satisfactory result (using the sources from Fig. \ref{fig_sources2}) is presented in Fig. \ref{fig_bench2}. Discrepancies up to 20\% are observed, with the largest disagreement in the ion temperatures (compare black versus green). The main reason for the disagreement in $T_{\rm i}$ is a different boundary condition used in the codes for $T_{\rm i}$, Eq. (\ref{eq_bc1}) in SOLF1D versus Eq. (\ref{eq_bc2}) in SOLPS
\begin{eqnarray}
5/2nkT_{\rm i}u_{\parallel}+1/2m_{\rm i}nu_{\parallel}^3+q_{\rm \parallel,i}+u_{\parallel}\delta p_{\rm i} = 3.5 nkT_{\rm i}c_{\rm s}, \label{eq_bc1} \\
3/2nkT_{\rm i}u_{\parallel}+q_{\rm \parallel,i} = 2.5 nkT_{\rm i}c_{\rm s}, \label{eq_bc2}
\end{eqnarray}
see sections \ref{sec_solf1d}-\ref{sec_solps}. Fig. \ref{fig_bench2} shows that if we use the same boundary condition in SOLF1D as the one defined in SOLPS, a perfect match of $T_{\rm i}$ is obtained (compare black versus red), however a disagreement in $n$ and $T_{\rm e}$ is still present.

Several checks were carried out to identify the cause of the remaining disagreement, including a test of the grid resolution (the SOLF1D solutions in Figs. \ref{fig_bench1}-\ref{fig_bench2} are spatially converged), a check of boundary conditions, a comparison for simplified cases reducing physics of the model, a test of inaccuracy due to the 2D numerical discretization in SOLPS, which will be described below.  

\subsection{Boundary conditions}
\begin{figure}[!h]
\centerline{\scalebox{0.7}{\includegraphics[clip]{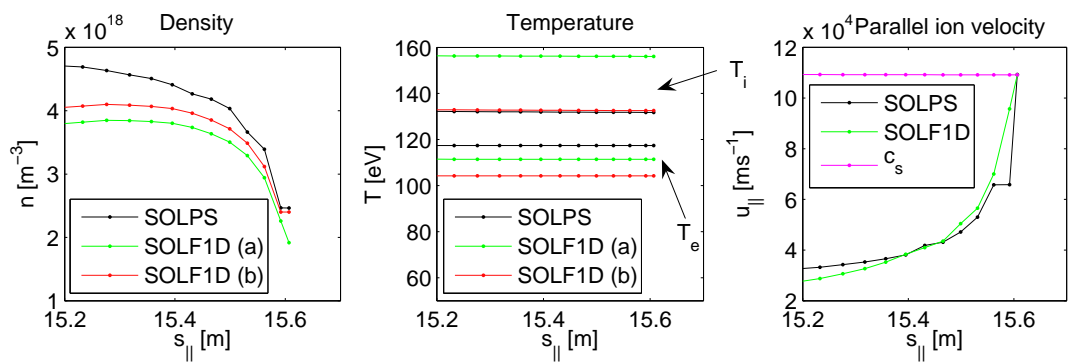}}}
\caption{Boundary conditions in the codes. (i) Plasma density close to the target from SOLPS (black), SOLF1D using the standard extrapolation boundary condition (case a) and SOLF1D using the zero gradient boundary condition (case b). (ii) Plasma temperatures close to the target shown as a comparison of SOLPS solution versus SOLF1D result using the standard SOLF1D boundary conditions (case a) or using boundary conditions identical to SOLPS (case b). (iii) Parallel ion velocity at the target follows the same boundary condition in both codes, $u_{\parallel}=c_{\rm s}$, and the solution is shown for the same temperature in the codes.}
\label{fig_other1}
\end{figure}

Fig. \ref{fig_other1} shows plasma quantities at the target with different boundary conditions used in the codes. As it is more complicated to change boundary conditions in SOLPS, the easiest way to benchmark the codes is to modify those in SOLF1D. Therefore, for further comparisons below, boundary conditions in SOLF1D are fixed to the SOLPS ones, i.e. the boundary condition for $T_{\rm i}$ complies Eq. \ref{eq_bc2} and the boundary condition for $n$ complies $\nabla_{\parallel}n=0$ at the target.

\subsection{Reduced model}

\begin{figure}[!h]
\centerline{\scalebox{0.7}{\includegraphics[clip]{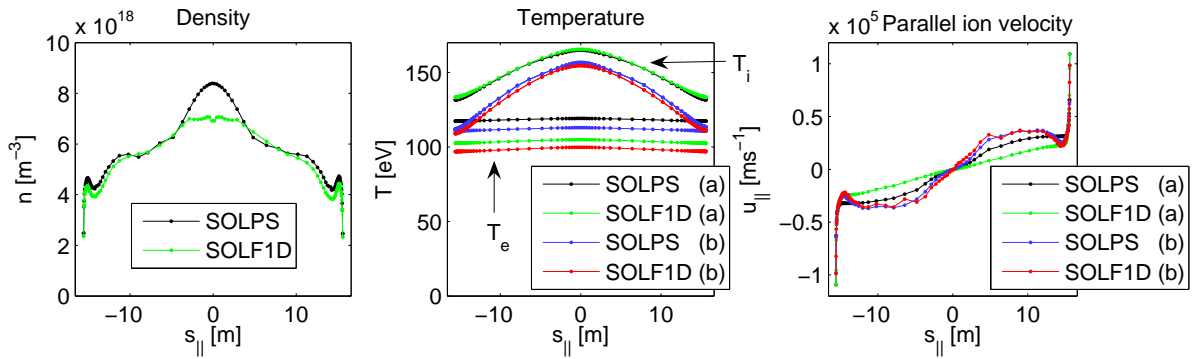}}}
\caption{A flux tube in the CD grid -- comparison of SOLF1D and SOLPS solutions for a reduced model. (i) Benchmark of density solvers only (left). (ii) Benchmark of temperature solvers only (middle) -- (a) full model, (b) reduced model neglecting viscous heating and electron-ion equipartition. (iii) Benchmark of velocity solvers only (right) -- (a) full model, (b) reduced model neglecting viscous fluxes.} \label{fig_bench3}
\end{figure}

For the same case (the flux tube in the CD geometry in Fig. \ref{fig_bench2}), Fig. \ref{fig_bench3} shows the best agreement we can achieve between the codes. Here, we use the same boundary conditions in SOLF1D as in SOLPS for all quantities and we reduce the equations of the model to identify possible problematic terms. We test the continuity, momentum and energy solvers one by one, i.e. the density shown in Fig. \ref{fig_bench3} is obtained by fixing the velocity and temperature in SOLF1D to the SOLPS profiles so that only the continuity solvers are tested, and similarly the temperature and velocity profiles in Fig. \ref{fig_bench3} are obtained with fixing the other quantities. (i) A comparison of the densities (left) shows a fair agreement at the targets, while in the upstream SOL, SOLF1D gives 18\% smaller density than SOLPS for the same particle source from Fig. \ref{fig_sources2} and the magnetic field from Fig. \ref{fig_scheme}. (ii) The ion temperatures agree well everywhere on the flux tube, while the flat electron temperature predicted in SOLF1D is approximately 11\% lower (case a). This trend remains even if we neglect the viscous heating and the energy exchange between electrons and ions (case b). (iii) The parallel ion velocities disagree by 35\% (case a), however a perfect match is obtained if the viscous flux is neglected (case b). Indeed, SOLPS takes into account an additional viscous flux driven by $\nabla_{\parallel}q$ which is not included in SOLF1D, see section \ref{sec_solps}. Based on this benchmark, it is considered to include such term in SOLF1D as well. Apart from the viscous term, no other differences in physics of the models have been identified. 

\subsection{Numerical discretization}
The persisting disagreement in the density and temperature profiles is likely caused by the numerical discretization. In Figs. \ref{fig_bench2} and \ref{fig_bench3}, the SOL is in the sheath-limited regime. This generally means that the particle transport, Eq. (\ref{eq_fvm}), is governed by the divergence of the flux, while in the high-recycling regime (Fig. \ref{fig_bench1}), it is the recycling source that dominates. The recycling source is calculated in EIRENE and it is treated in the same way in both codes as a net source. The origin of the disagreement in Fig. \ref{fig_bench3} therefore comes most likely from the way the flux divergence is calculated and how this term is discretized on the grid. This problem is masked in Fig. \ref{fig_bench1} in high-recycling conditions. 

\begin{figure}[!h]
\centerline{\scalebox{0.65}{\includegraphics[clip]{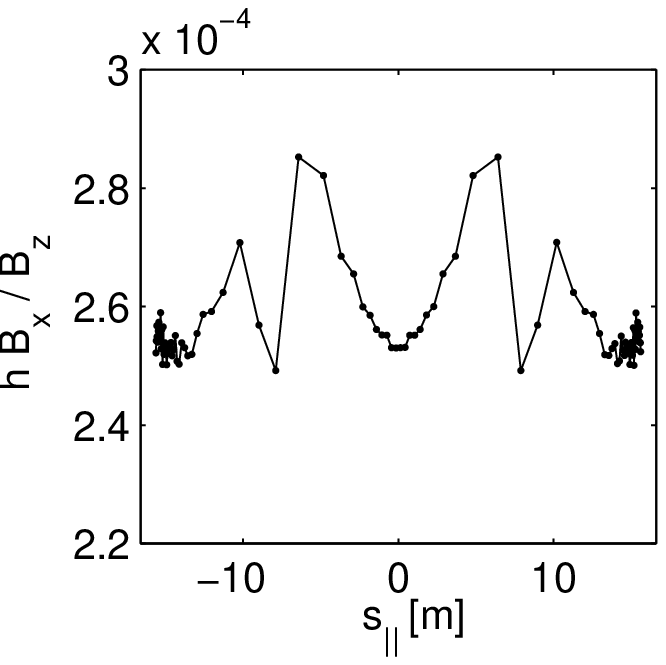}}
\scalebox{0.7}{\includegraphics[clip]{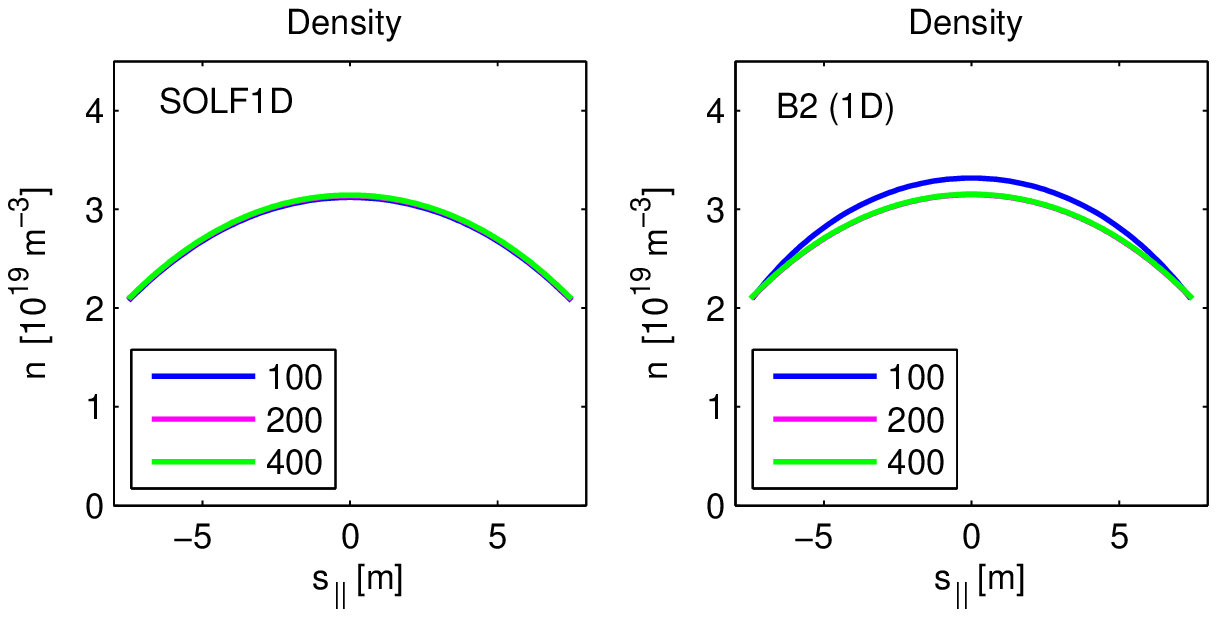}}}
\caption{(i) Poloidal magnetic flux along the flux tube expressed as $hB_x/B_z$ where $h$ is the radial width of the flux tube (left). (ii) Test of spatial convergence of the SOLF1D continuity solver comparing solutions for different number of parallel grid points (middle). (iii) Spatial convergence of SOLPS (right).} \label{fig_other2}
\end{figure}

\begin{figure}[!h]
\centerline{\scalebox{0.7}{\includegraphics[clip]{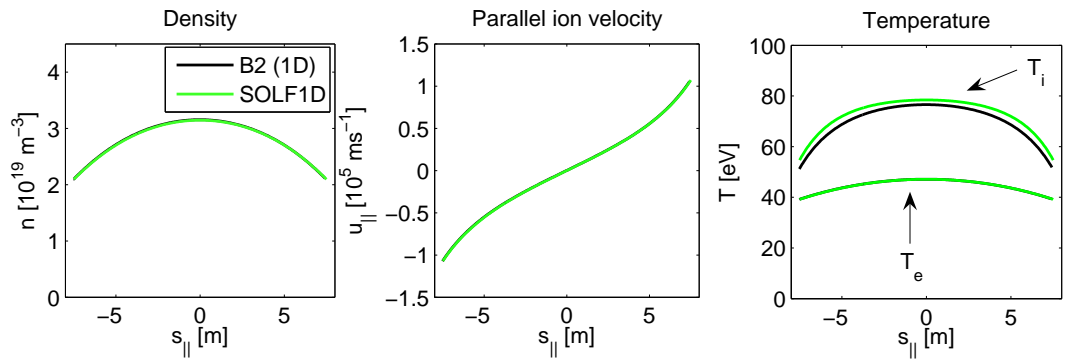}}}
\caption{Comparison of solutions from SOLF1D and 1D version of SOLPS. The plasma density and parallel ion velocity (left and middle) are compared independently of the temperature solver with temperatures fixed to 60 eV. The electron and ion temperatures for a reduced model neglecting ion viscous heating are shown on the right.} \label{fig_other3}
\end{figure}

As far as the discretization of the magnetic topology is concerned, it was shown in section \ref{sec_solps} that the SOLPS equations are identical to the 1D equations solved in SOLF1D, if the condition of constant poloidal magnetic flux on the discretized flux surface is fulfilled. Fig. \ref{fig_other2} (left) indicates that the poloidal magnetic flux $hB_x/B_z$ is constant in the divertor, however, it oscillates by approximately 12\% above the average value around X points. Similar level of discrepancy can be expected in the comparison of SOLPS/SOLF1D solutions. We can eliminate this numerical error by comparing SOLF1D with a 1D version of SOLPS, assuming $B={\rm const}$ along the flux tube. Such comparison has been done previously in \cite{Eva2}. Identical solutions were obtained, see Fig. \ref{fig_other2} (middle and right), but only when the grid resolution in SOLPS was doubled from 100 to 200 grid points, and approximately 6\% error in the density solution in SOLPS was detected for 100 grid points. Note that SOLPS does not typically run on more than 100 poloidal cells, which is the case also here. Fig. \ref{fig_other3} compares solutions from SOLF1D and 1D SOLPS directly for 400 grid points. At high grid resolution, the continuity and momentum solvers are identical and the electron temperatures agree as well. 

Beside an inaccuracy in the discrete magnetic topology, Fig. \ref{fig_other2} shows an inaccuracy of the discretized equations which can be suppressed by refining the grid. The different sensitivity of the codes to the grid resolution can be explained by different numerical schemes used in the codes to treat the flux divergence and the different methods to solve the discretized system of equations. The numerical technique of SOLF1D is described in the appendix. SOLPS is based on the Patankar hybrid scheme and uses the velocity-pressure coupling approach. While SOLF1D uses a staggered grid, SOLPS stores $u_{\parallel}$ at the cell centres and uses an interpolation method to calculate fluxes at the cell faces. Indeed, differences (largest around the stagnation points) have been seen when switching from the cell-faced to cell-centred version of SOLPS \cite{DC2}. A numerical error can also result from the evaluation of the flux divergence as the fluxes in and out of the flux tube have very similar and large values. Tab. \ref{tab0} briefly summarizes this section.

\begin{table}[!h]
\begin{center}
\begin{tabular}{l|cccc|p{10cm}}
 & $n$ & $u_{\parallel}$ & $T_{\rm e}$ & $T_{\rm i}$ \\
\hline
1. & \ding{55} & \ding{55} & \ding{55} & \ding{55} & standard SOLF1D model \\
2. & \ding{55} & \ding{55} & \ding{55} & \ding{51} & SOLF1D model using BC as defined in SOLPS \\
3. & \ding{55} & \ding{51} & \ding{55} & \ding{51} & models assuming zero viscous flux \\
4. & \ding{51} & \ding{51} & \ding{51} & \ding{51} & comparison of models in 1D with high grid resolution \\
\hline
\end{tabular}
\caption{Steps towards agreement in quantities solved by SOLF1D and SOLPS codes. 1. The standard SOLF1D model results in up to 20\% different solution with respect to SOLPS. 2. The SOLF1D model using boundary conditions as defined in SOLPS results in agreement in $T_{\rm i}$ (the reason for the discrepancy -- the kinetic heat flux is not included in the SOLPS boundary conditions for $T_{\rm i}$). 3. Models assuming zero viscous flux agree in $u_{\parallel}$ (the reason for the discrepancy -- SOLPS takes into account an additional heat flux driven viscosity not included in SOLF1D). 4. A comparison in 1D with high grid resolution results in agreement in $n_{\rm e}$ and $T_{\rm e}$ (the reason for the discrepancy -- the poloidal magnetic flux in SOLPS varies along the flux tube and/or the SOLPS solution is spatially converged for $N\geq 200$ grid points only).}
\label{tab0}
\end{center}
\end{table}


\section{1D effects of long-legged divertor}

\subsection{Analysis of sources on a flux tube}
Power losses and particle sources in the SOL are affected by the interaction of plasma with neutral species, which is typically dominant in the divertor. The power and particle SOL balance depends on the collisionality regime where the recycling at the targets competes with the cross-field transport dominant in the upstream SOL. If the target fluxes are of interest, it is useful to analyze sources in the SOL as they drive the flow and their integral along the SOL gives the target flux directly. In Figs. \ref{fig_sources1} and \ref{fig_sources2}, sources of particles, momentum and energy are shown on a flux tube in CD and SXD, separated as recycling ($S_{\rm coll}$) and cross-field ($S_{\perp}$) sources $S=S_{\perp}+S_{\rm coll}$. In these figures, both conduction-limited and sheath-limited SOL display dominant particle source at the targets due to ionization ($S^n_{\rm coll}$), compared to the cross-field transport source ($S^n_{\perp}$) which is strong around the stagnation point and below X points. While the magnitude of $S^n_{\perp}$ does not change much between CD and SXD, $S^n_{\rm coll}$ is a factor of 3 larger (note the different scale for $S^n_{\perp}$ and $S^n_{\rm coll}$). The energy source, on the other hand, is in both cases dominated by the radial transport, therefore enhanced electron cooling in SXD does not have a big impact on the power balance in the investigated regime (attached plasma). A detailed analysis of sources in the two topologies is given in Tab. \ref{tab1}, where integral sources along the flux tube (representing target fluxes) are calculated, clearly identifying the dominant terms and the terms that change the most with the expanded divertor leg.

\begin{table}[!h]
\begin{center}
\begin{tabular}{lrrr}
    & CD & SXD \\
\hline
$\int S^n {\rm d}s_{\parallel}$ [m$^{-2}$s$^{-1}$] & $5.0 \times 10^{23}$ & $1.8 \times 10^{24}$ & 3.5 \\
$\int S^n_{\perp} {\rm d}s_{\parallel}$ [m$^{-2}$s$^{-1}$] & $-9.1 \times 10^{21}$ & $-3.5 \times 10^{23}$ & 38.5 \\
$\int S^n_{\rm coll} {\rm d}s_{\parallel}$ [m$^{-2}$s$^{-1}$] & $5.1 \times 10^{23}$ & $2.1 \times 10^{24}$ & 4.2 \\
\hline
$\int S^E {\rm d}s_{\parallel}$ [${\rm kg \, s}^{-3}$] & $4.6 \times 10^{7}$ & $6.9 \times 10^{7}$ & 1.5 \\
$\int S^E_{\rm \perp,e} {\rm d}s_{\parallel}$ [${\rm kg \, s}^{-3}$] & $3.5 \times 10^{7}$ & $6.2 \times 10^{7}$ & 1.8 \\
$\int S^E_{\rm \perp,i} {\rm d}s_{\parallel}$ [${\rm kg \, s}^{-3}$] & $1.7 \times 10^{7}$ & $1.7 \times 10^{7}$ & 1.0 \\
$\int S^E_{\rm coll,e} {\rm d}s_{\parallel}$ [${\rm kg \, s}^{-3}$] & $-2.2 \times 10^{6}$ & $-8.7 \times 10^{6}$ & 4.0 \\
$\int S^E_{\rm coll,i} {\rm d}s_{\parallel}$ [${\rm kg \, s}^{-3}$] & $-3.7 \times 10^{6}$ & $-2.3 \times 10^{6}$ & 0.6 \\
\hline
$B_{\rm t}\int S^n/B {\rm d}s_{\parallel}$ [m$^{-2}$s$^{-1}$] & $5.0 \times 10^{23}$ & $1.6 \times 10^{24}$ & 3.2 \\
$B_{\rm t}\int S^E/B {\rm d}s_{\parallel}$ [${\rm kg \, s}^{-3}$] & $8.0 \times 10^{7}$ & $4.7 \times 10^{7}$ & 0.6 \\
$B_{\rm t}$ [T] & $0.8$ & $0.4$ & 0.5 \\
\hline
$Q_{\parallel}$ [${\rm kg \, s}^{-3}$] & $3.9 \times 10^{7}$ & $2.5 \times 10^{7}$ & 0.6 \\
$Q_{\rm \parallel,e}$ [${\rm kg \, s}^{-3}$] & $2.0 \times 10^{7}$ & $1.7 \times 10^{7}$ & 0.9 \\
$Q_{\rm \parallel,i}$ [${\rm kg \, s}^{-3}$] & $1.9 \times 10^{7}$ & $8.3 \times 10^{6}$ & 0.4 \\
$\Gamma_{\parallel}$ [m$^{-2}$s$^{-1}$] & $2.2 \times 10^{23}$ & $8.3 \times 10^{23}$ & 3.7 \\
\hline
\end{tabular}
\caption{(top) Integral particle and energy sources on a flux tube of the CD and SXD configurations (the flux tube with maximum target energy load connecting the outer targets). $S^n$ is determined by $S^n_{\rm coll}$, which increases $4\times$ with increased $L_{\parallel}$. Although $S^E_{\rm coll,e}$ changes $4\times$, $S^E$ is mainly determined by $S^E_{\perp}$, which has weaker dependence on $L_{\parallel}$, as well as $S^E_{\rm coll,i}$ which changes by factor of 0.6. Note that the flux tube on the outer side represents well the 2D picture, as the total change of particles by collisional processes in the entire grid volume is by factor of 3, the electron energy is reduced by factor of 3 and the change in the ion energy is of factor of 0.6. (bottom) As a consequence of mass and energy conservation on the flux tube, the integral sources represent the target particle and energy fluxes $\Gamma_{\parallel}$ and $Q_{\parallel}$ through expressions $B_{\rm t}\int S^n/B {\rm d}s_{\parallel}=2\Gamma_{\parallel}$, $B_{\rm t}\int S^E/B {\rm d}s_{\parallel}=2Q_{\parallel}$ where $B_{\rm t}$ is the magnetic field at the target. This is found by the integration of Eq. (\ref{eq_continuity}), (\ref{eq_energy1}) and (\ref{eq_energy2}).
}
\label{tab1}
\end{center}
\end{table}

\subsection{Stretching the flux tube}
\begin{figure}[!h]
\centerline{\scalebox{0.3588}{\includegraphics[clip]{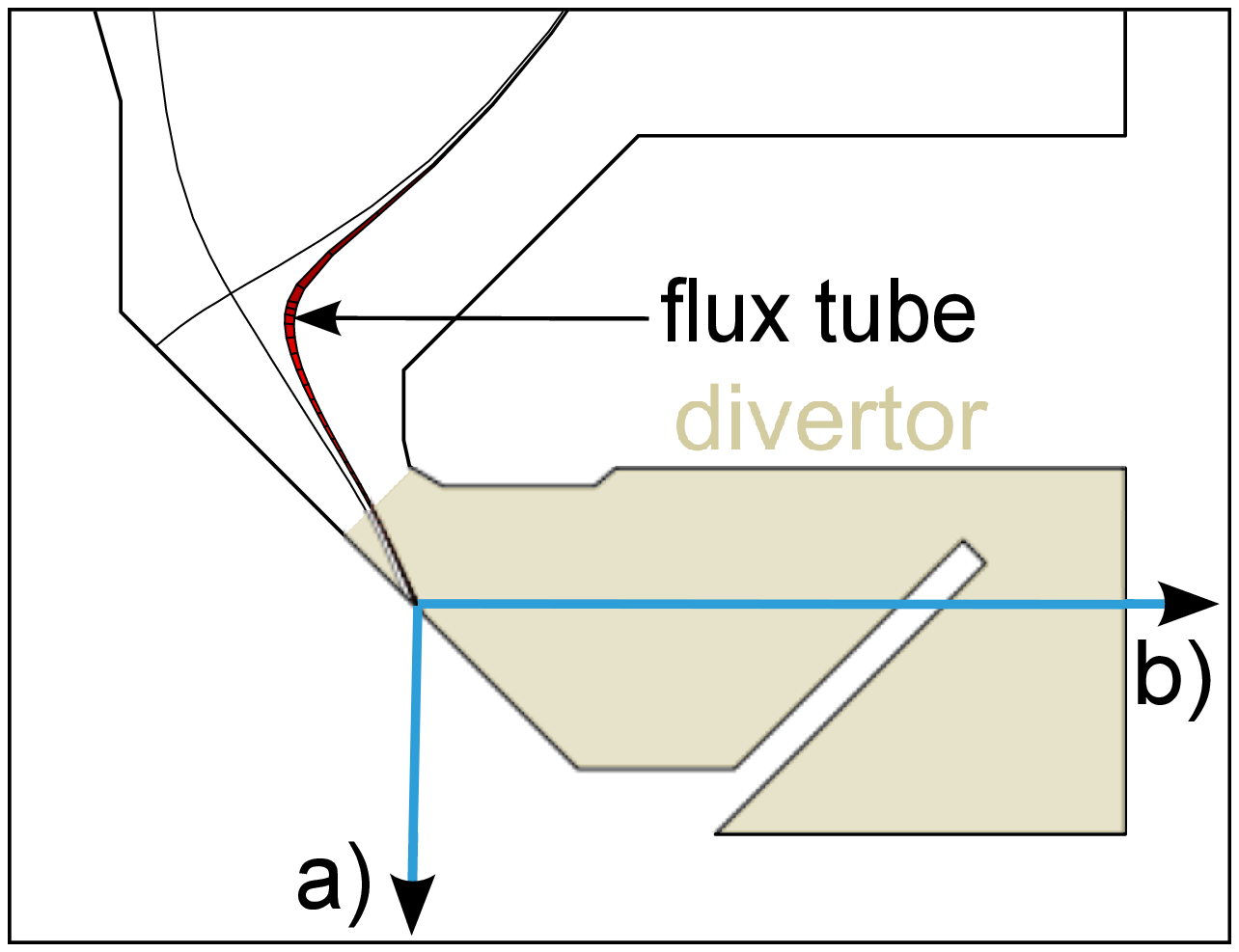}}
\scalebox{0.7176}{\includegraphics[clip]{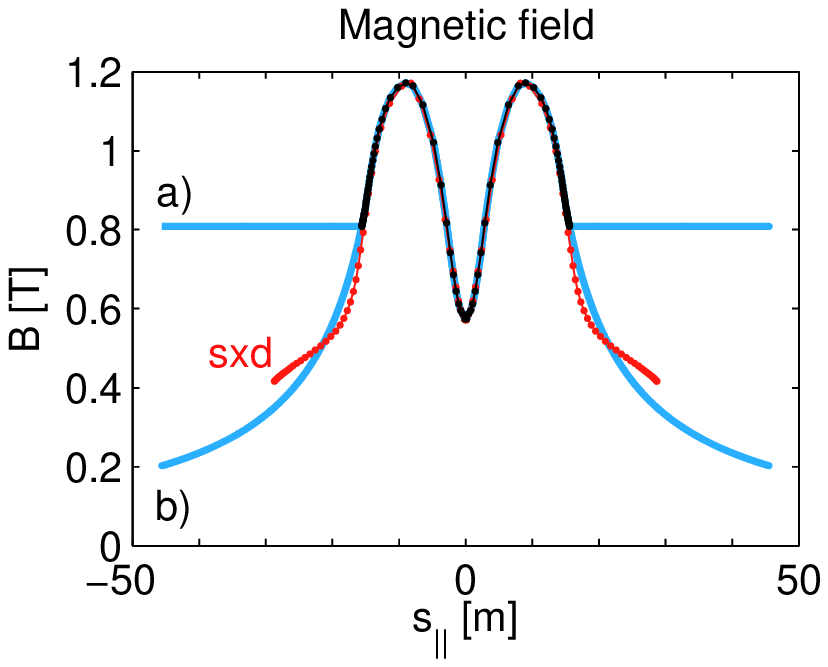}}
}
\caption{The scheme on the left shows a flux tube from SOLPS and different ways of stretching this flux tube in the divertor in 1D: (a) a vertical stretching (the magnetic field is constant in the divertor), (b) a radial stretching (the magnetic field decreases as $B\propto 1/R$). The graph on the right shows the magnetic field considered in the 1D code.}
\label{fig_stretch}
\end{figure}

As the particle and power balance on a flux tube in the outer SOL from Fig. \ref{fig_sources2} (Tab. \ref{tab1}) seems to represent well the balance in the entire outer SOL (see a comment in Tab. \ref{tab1}), we use the 1D approach to estimate the change in target parameters that occurs due to stretching the flux tube in the divertor. We start with the short divertor configuration (Fig. \ref{fig_bench2}) and expand the flux tube in two directions (Fig. \ref{fig_stretch} left) 
labelled as (a) and (b). The magnetic field along the flux tube corresponding to these two directions is plotted in Fig. \ref{fig_stretch} on the right (blue) and compared to the original CD case from Fig. \ref{fig_bench2} (black) and the SXD case from Fig. \ref{fig_bench1} (red). 

Particle and energy sources in the expanded divertor are not self-consistently modelled in 1D, as it would only give a rough approximation of the 2D transport. Instead, they are approximated by a parametric dependence on the divertor length $L_{\rm div}$, using sources calculated by EIRENE for CD at $L_{\parallel}\approx 15$ m. The radial and recycling sources are treated separately. We assume that $S_{\perp}$, which is largest in the upstream SOL and below the X points, does not change during stretching and it is zero in the newly created region. Further, we define a divertor region (Fig. \ref{fig_stretch} left -- the shaded region between the nose of the baffle and target) where we replace $S_{\rm coll}$ by its average value in this region $\overline{S}_{\rm coll}$. This does not change the target flux, but simplifies the treatment of sources during stretching of the divertor leg. 
For the recycling source, we employ two methods: (1) While expanding the flux tube, we assume that $\int S_{\rm coll}$ is constant and $S_{\rm coll}$ in the divertor is replaced by its average value $\overline{S}_{\rm coll}$. (2) We assume that the uniformly distributed source in the divertor $\overline{S}_{\rm coll}$ is kept constant, i.e. the integrated source (or sink) $\int S_{\rm coll}$ grows with $L_{\rm div}$. Method (1) assumes that the sources due to plasma-neutral interaction do not increase for increased divertor length, and is only used to separate the effect of toroidal magnetic flux expansion. Method (2) is more realistic as it incorporates plasma-neutral cooling and increased ionization source for increased $L_{\rm div}$. A comparison with sources from SOLPS for SXD (Tab. \ref{tab1}) shows the suggested approximation of the sources is reasonable.

\subsection{Target parameters as function of $L_{\parallel}$}

\begin{figure}[!h]
\centerline{\scalebox{0.63}{\includegraphics[clip]{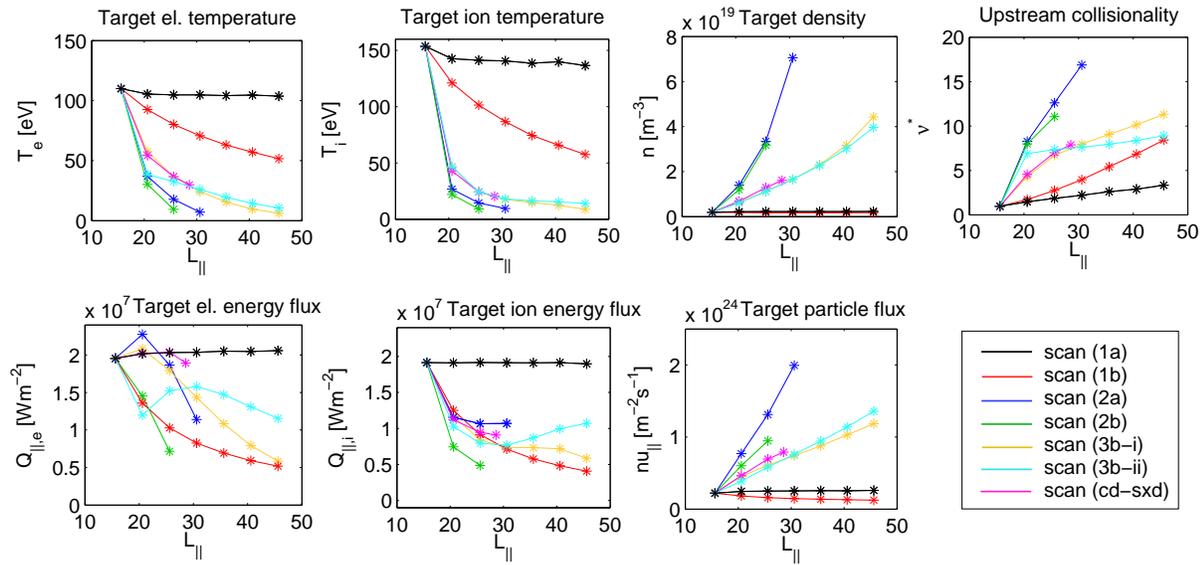}}}
\caption{Parameters at the target calculated by SOLF1D as functions of the connection length, based on scans involving stretching of the divertor region as described in the text. (1a) Constant integral sources in the flux tube (i.e. sources in the divertor are reduced proportionally to the increased divertor length), constant magnetic field in the divertor (see Fig. \ref{fig_stretch} a). (1b) Constant integral sources, varying magnetic field (Fig. \ref{fig_stretch} b). (2a) Sources in the divertor multiplied by the divertor length, constant magnetic field in the divertor. (2b) Sources in the divertor multiplied by the divertor length, varying magnetic field. (3b-i) Sources in the divertor varied as $S\propto L_{\rm div}^{\alpha}$, the magnetic field as in case (b). (3b-ii) Sources in the divertor varied as $S\propto L_{\rm div}$, the magnetic field as in case (b). (cd-sxd) Sources in the divertor varied as $S\propto L_{\rm div}^{\alpha}$, the magnetic field as in SXD case. The cases (3b) and (cd-sxd) will be defined later in section \ref{better_app}.}
\label{fig_scan1d}
\end{figure}

The different treatments of the sources in the flux tube, (1) versus (2) described above, and different directions of stretching the flux tube, (a) versus (b) from Fig. \ref{fig_stretch}, give a combination of 4 scans, each shown in Fig. \ref{fig_scan1d}.

(1a) -- black -- For the case of the vertical stretching (no additional flux expansion in the divertor), the target quantities do not change and both the particle and energy fluxes remain the same with increasing $L_{\rm div}$ ($B_{\rm t}\int S^n /B{\rm d}s_{\parallel}$ and $B_{\rm t}\int S^E /B{\rm d}s_{\parallel}$ do not change with constant magnetic field in the divertor). 
(1b) -- red -- The radial stretching (the magnetic field drops in the divertor as $B\propto 1/R$) leads already to a large reduction of temperatures and energy fluxes. This case shows the effect of magnetic flux expansion solely. 
(2a) -- blue -- With more recycling in the divertor, i.e. larger particle source due to ionization and larger cooling (the recycling sources scale with the divertor volume), the temperature drop is much steeper even without magnetic flux expansion, compared to the case (1a), however, we see less effect on the energy fluxes than from the flux expansion alone in the previous case. 
(2b) -- green -- The most beneficial, in terms of $Q_{\parallel}$ reduction, is the combination of both effects, which results in a substantial drop of temperature and a moderate drop of energy fluxes, however, also in higher collisionality and increased density and particle flux under attached conditions. The approximation (2b) predicts a transition to detachment at approximately $L_{\parallel}\approx 25$ m, i.e. SXD at $L_{\parallel}\approx 28$ m would be detached. 

\subsection{Approximation of sources in 1D code}
\label{better_app}

Although the approximation above for the sources captures well the trends, it does not recover exactly the SOLF1D/SOLPS simulation for SXD in the 1D scan. This is because too strong a dependence of the recycling sources on $L_{\rm div}$ was assumed. In SXD, where $L_{\rm div}$ is $8\times$, longer, the ionization source and cooling in the divertor is only $4\times$ stronger, while the approximation (2) assumes a linear relationship. Therefore, the methods (1) and (2) are extreme cases and the actual SXD SOLPS simulation is somewhere in between.


If we want to approximate the sources to describe the transition from CD to SXD quantitatively, we have to pay attention to those components that are dominant. From Tab. \ref{tab1}, the key source terms to drive fluxes are $\int S^n_{\rm coll}$ for particles and $\int S^E_{\perp}$ for energy, although the change of $\int S^E_{\rm coll,e}$ can not be neglected as $\int S^E_{\rm coll,e}$ grows with  collisionality and $L_{\parallel}$ and will be important at larger $L_{\parallel}$. In addition, Tab. \ref{tab1} shows an increase of the radial electron energy source in SXD. Because this source dominates over the collisional cooling occurring in the divertor in both CD and SXD, this has an effect on the evaluation of the target energy flux. Comparison of Fig. \ref{fig_sources1} and \ref{fig_sources2} shows that the sink of electron and ion energy below the X point into the private flux region is smaller in SXD and the amplitude of the electron energy source is larger. The increase of the radial energy source in SXD is consistent with stronger parallel electron energy transport governed by conduction, induced by steeper temperature gradients in SXD ($S_{\rm \perp,e}^{E}\approx S_{\rm \parallel,e}^{E}\approx -\kappa_{\rm e}kT_{\rm e}/L_{\nabla T_{\rm e}}$, $L_{\nabla T_{\rm e}}^{\rm SXD}<L_{\nabla T_{\rm e}}^{\rm CD}$), i.e. the same flux comes radially into the flux tube from the main plasma, but there is less flux leaving the flux tube in SXD as a consequence of a stronger parallel loss, therefore the source is stronger.

Thus, based on the results from SOLPS in Tab. \ref{tab1}, we define a third method (3), an intermediate case, so that we fit the CD and SXD cases accurately and extrapolate to larger $L_{\parallel}$. 
We assume that $\int S_{\rm coll}^n$ and $\int S_{\rm coll,e}^E$ are such functions of the divertor length $L_{\rm div}$, that both CD and SXD cases in Tab. \ref{tab1} are matched. We keep $\int S_{\rm i}^E$ unchanged, but the amplitude of $S_{\rm \perp,e}^E$ increases with $L_{\rm div}$ as in SXD. This approximation is then used to predict target parameters for larger $L_{\rm div}$ beyond SXD. To interpolate between the CD and SXD cases, we assume a power dependence $\int S \propto L_{\rm div}^{\gamma}$ (i), and we also examine a linear dependence 
(ii). A complete list of scans is given in Tab. \ref{tab2}.

\begin{table}[!h]
\begin{center}
\begin{tabular}{lll}
\hline
scan (1a)& $\mathbb{S}_{\rm div}=\mathbb{S}_0$ & $B_{\rm div}={\rm const}$ \\
scan (1b)& $\mathbb{S}_{\rm div}=\mathbb{S}_0$ & $B_{\rm div}\propto 1/R$ \\
scan (2a)& $\mathbb{S}_{\rm div}=\mathbb{S}_0L_{\rm div}/L_0$ & $B_{\rm div}={\rm const}$ \\
scan (2b)& $\mathbb{S}_{\rm div}=\mathbb{S}_0L_{\rm div}/L_0$ & $B_{\rm div}\propto 1/R$ \\
scan (3a-i)& $\mathbb{S}_{\rm div}=\mathbb{S}_0(L_{\rm div}/L_0)^{\gamma}$ & $B_{\rm div}={\rm const}$ \\
scan (3b-i)& $\mathbb{S}_{\rm div}=\mathbb{S}_0(L_{\rm div}/L_0)^{\gamma}$ & $B_{\rm div}\propto 1/R$ \\
scan (3b-ii)& $\mathbb{S}_{\rm div}=\mathbb{S}_0\gamma L_{\rm div}/L_0$ & $B_{\rm div}\propto 1/R$ \\
scan (cd-sxd)& $\mathbb{S}_{\rm div}=\mathbb{S}_0(L_{\rm div}/L_0)^{\gamma}$ & $B_{\rm div}$ as in SXD \\
\hline
\end{tabular}
\caption{The definition of cases considered in the analysis -- the treatment of the sources and the magnetic field in the expanded divertor. $\mathbb{S}_{\rm div}=\int S_{\rm coll}$ is the integral source in the divertor region of the length $L_{\rm div}$ and $\mathbb{S}_0$ is its reference value for the conventional divertor of the length $L_0$. The coefficients $\gamma$ in the cases (3) and (cd-sxd) are chosen from fitting the SXD case.}
\label{tab2}
\end{center}
\end{table}

Fig. \ref{fig_scan1d} shows the intermediate case (3), in which the sources best fit the CD and SXD simulations, in addition to the previous cases (1) and (2). On the other hand method (2) leaves space for additional cooling, e.g. via impurity radiation which is not taken into account here. For the case (3) in Fig. \ref{fig_scan1d}, only radial stretching (b) is considered and 2 different fits, (i) and (ii), are assumed. The magenta case shown as (cd-sxd) is the direct interpolation between CD and SXD, using the dependence (i). This case almost copies the case (3b-i) where the divertor leg is stretched radially (the $B$ dependence is shown by the blue curve on right hand side of Fig. \ref{fig_stretch}), with the only difference of assuming directly the magnetic field and geometry of SXD (the red curve in Fig. \ref{fig_stretch}). The last magenta point at $L_{\parallel}\approx 28$ m therefore directly corresponds to targets parameters found in SXD by SOLF1D/SOLPS, while the CD case is at $L_{\parallel}\approx 15$ m.

Cases (3b-i) and (3b-ii) represent an extrapolation from SXD to a divertor at larger radius, with different functions used to interpolate between CD and SXD. These cases lie between the limiting cases (1b) and (2b), and for the initial condition of $T_{\rm e,sep}$ both lead to a reduction of $T_{\rm e}$ approximately as $T_{\rm e}\propto L_{\parallel}^{-2.6}$. At $L_{\parallel}\approx 28$ m, $T_{\rm e}$ is still above the detachment limit of 5 eV, for which $L_{\parallel}\approx 45$ m is needed. At $L_{\parallel}<45$ m, the reduction of the energy fluxes achieved in cases (3b-i) and (3b-ii) does not exceed the reduction caused by the magnetic flux expansion solely in case (1b). It is interesting to see that $Q_{\rm \parallel,e}$ at $L_{\parallel}<30$ m is unchanged. The reason for this is, that in spite of stronger volumetric power losses with increased $L_{\parallel}$, reducing the target energy flux, the parallel electron heat flux, governed mainly by the source due to the cross-field transport, is enhanced as well (stronger $\nabla_{\parallel}T_{\rm e}$ at increased $L_{\parallel}$). These two effects compete and the collisional cooling starts to show a strong effect only at large $L_{\parallel}$ ($L_{\parallel}>45$ m) in this density regime, as shown in (3b-i). Note that different extrapolations (i) or (ii) lead to similar $n$ and $T$ in the divertor, while the prediction of $Q_{\parallel}$ for large $L_{\parallel}$ is more sensitive to the dependence of the collisional source on $L_{\parallel}$.

\subsection{Comparison with two-point model}
\label{2p}

\begin{figure}[!h]
\centerline{\scalebox{0.63}{\includegraphics[clip]{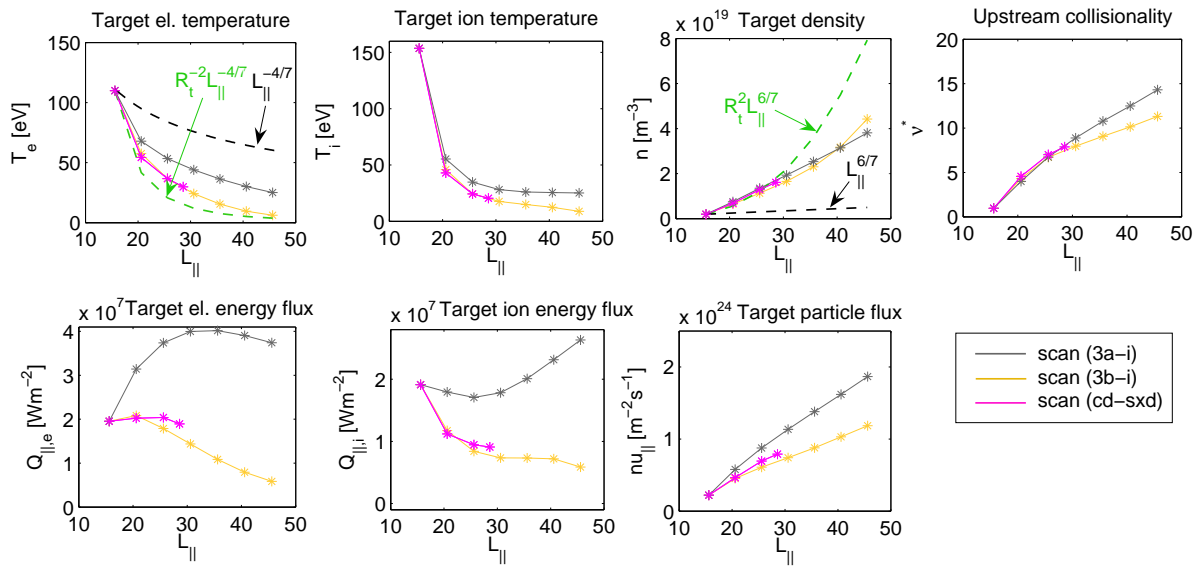}}}
\caption{Parameters at the target calculated by SOLF1D compared to the two-point model. (3a-i) Constant magnetic field in the divertor, sources in the divertor vary as $S\propto L_{\rm div}^{\alpha}$ and match the sources in CD and SXD at $L_{||}\approx 15$ m and $L_{||}\approx 28$ m. (3b-i) As in Fig. \ref{fig_scan1d}, i.e. the magnetic field in the divertor drops as $B\propto 1/R$, sources in the divertor go as $S\propto L_{\rm div}^{\alpha}$. (cd-sxd) As in Fig. \ref{fig_scan1d}, the magnetic field and sources change as in SXD, therefore plasma parameters at $L_{||}\approx 15$ m coincide with the CD values and at $L_{||}\approx 28$ m with the SXD values. For comparison, the two-point model without/with $R_{\rm t}$ dependence is shown in black/green.}
\label{fig_2p}
\end{figure}

In Fig. \ref{fig_2p}, the reduction of the electron temperature and the increase of the density at the target are compared with a two-point model prediction. As in Fig. \ref{fig_scan1d}, a scan using the SXD magnetic field is shown (magenta) with an extrapolation that extends the divertor even further radially (3b-i, yellow). In addition, a scan assuming the vertical stretching instead of the radial one is plotted (3a-i, grey). The case (3a-i), where $T_{\rm e}$ falls as $T_{\rm e}\propto L_{\parallel}^{-1.3}$, shows steeper temperature drop than predicted by the two-point model for this case ($T_{\rm e}\propto L_{\parallel}^{-4/7}$ and $n_{\rm e}\propto L_{\parallel}^{6/7}$). The cases (cd-sxd) and (3b-i), on the contrary, show weaker dependence than a modified two-point model taking into account the dependence on the target radius $R_{\rm t}$ ($T_{\rm e}\propto R_{\rm t}^{-2}L_{\parallel}^{-4/7}$ and $n_{\rm e}\propto R_{\rm t}^2L_{\parallel}^{6/7}$, see \cite{Petrie}). The two-point model predicts approximately $2\times$ smaller $T_{\rm e}$ in SXD (at $L_{||}\approx 28$ m) than SOLF1D/SOLPS.

\subsection{Parallel profiles for different divertor lengths}

\begin{figure}[!h]
\centerline{\scalebox{0.63}{\includegraphics[clip]{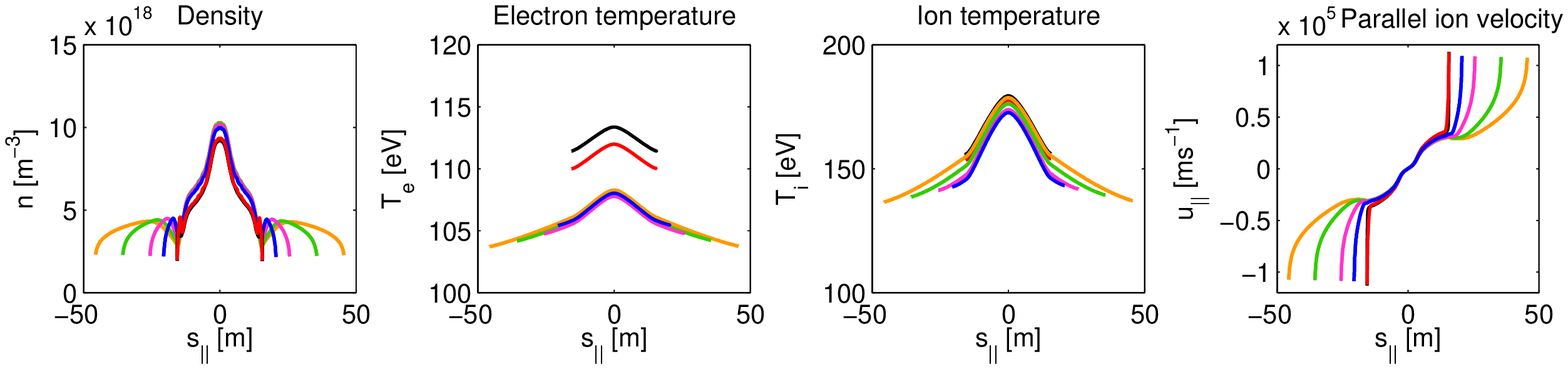}}}
\centerline{\scalebox{0.63}{\includegraphics[clip]{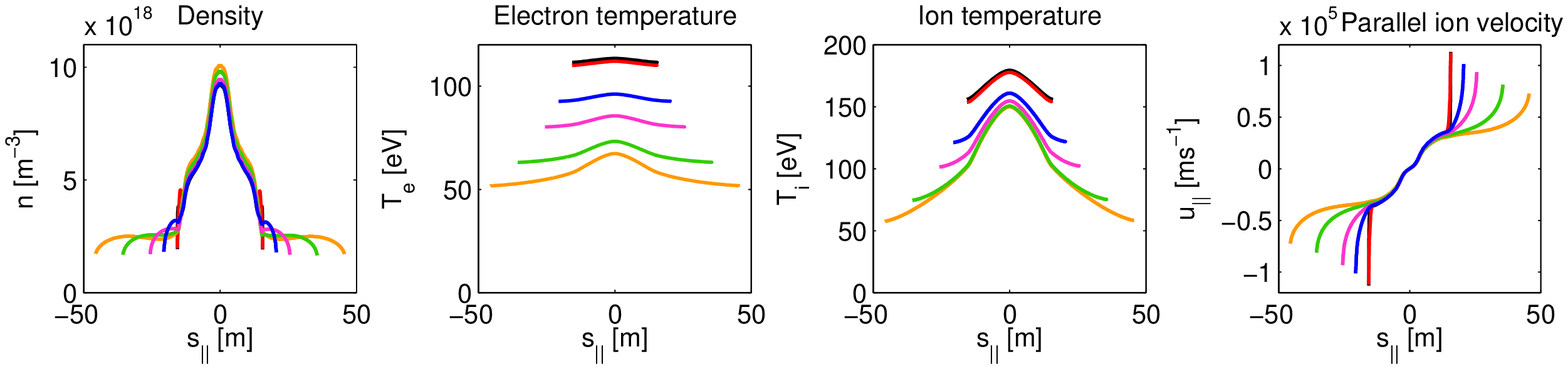}}}
\centerline{\scalebox{0.63}{\includegraphics[clip]{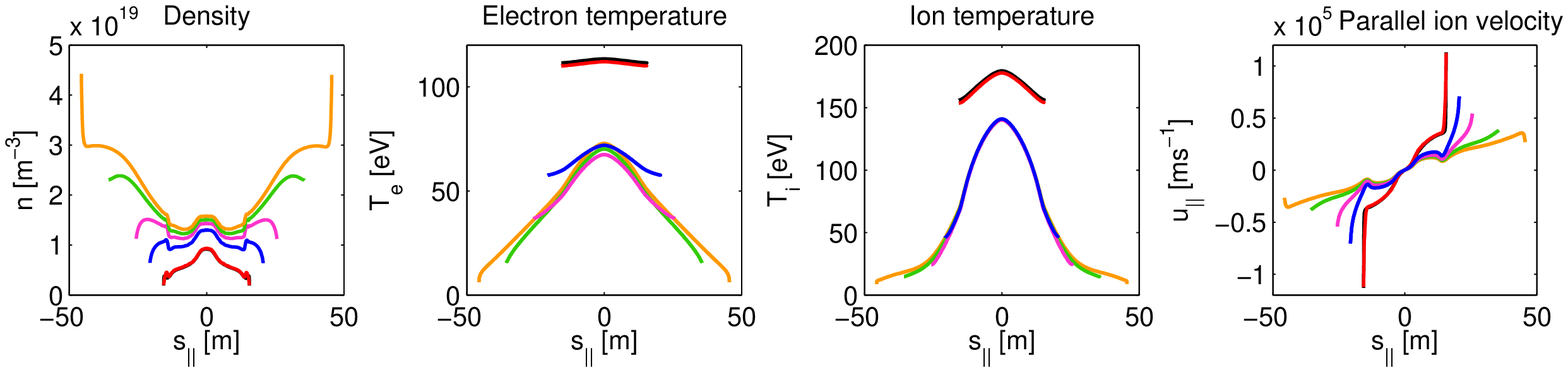}}}
\centerline{\scalebox{0.63}{\includegraphics[clip]{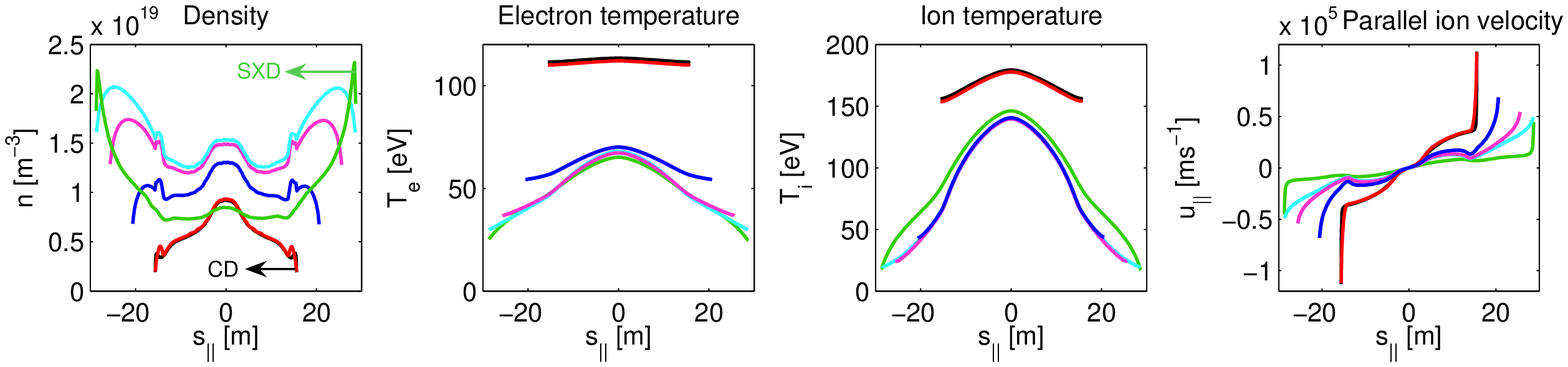}}}
\caption{Parallel profiles of plasma quantities on a flux tube between upper and lower outer targets for an expanded divertor leg: (i) Case (1a) from Fig. \ref{fig_scan1d}. (ii) Case (1b) from Fig. \ref{fig_scan1d} (increasing flux expansion). (iii) Case (3b-i) from Fig. \ref{fig_scan1d} (increasing flux expansion and the strength of recycling sources). In black, profiles for the CD geometry (as in Fig. \ref{fig_bench2}). In red, a solution for the CD geometry with the recycling sources at the target replaced uniformly by the average value in the divertor. In blue, magenta, green and orange, solutions for the expanded divertor. (iv) Case (cd-sxd) from Fig. \ref{fig_scan1d} (changing flux expansion and the strength of recycling sources as in SXD). In black, profiles for the CD geometry (from Fig. \ref{fig_bench2}). In blue, magenta and cyan, profiles for the expanded divertor. The last profile with $L_{||}=28$ m coincides at the targets with the solution for the SXD geometry (from Fig. \ref{fig_bench1}) shown in green. 
}
\label{fig_profiles1d}
\end{figure}

Fig. \ref{fig_profiles1d} shows parallel profiles for the scans from Fig. \ref{fig_scan1d}. In the first row (1a), the SOL is in the sheath-limited regime and no reduction of temperatures is observed. In the second row (1b), the target temperatures are reduced through the magnetic flux expansion (at the target $Q_{\parallel}\propto nkTc_{\rm s}$), but the electron temperature profiles remain flat (small parallel heat conductivity at large $T_{\rm e}$). In the third row (3b-i), the conduction-limited SOL is accessed by increased plasma-neutral cooling in the divertor accompanied by a drop of the plasma temperature at the target and an increase of the density. The last case displays profiles for the case (cd-sxd) from Fig. \ref{fig_scan1d} where the target parameters at $L_{\parallel}\approx 28$ m (cyan) coincide directly with the solution for SXD from Fig. \ref{fig_bench1} (green).

\subsection{Scan at lower temperature}

\begin{figure}[!h]
\centerline{\scalebox{0.63}{\includegraphics[clip]{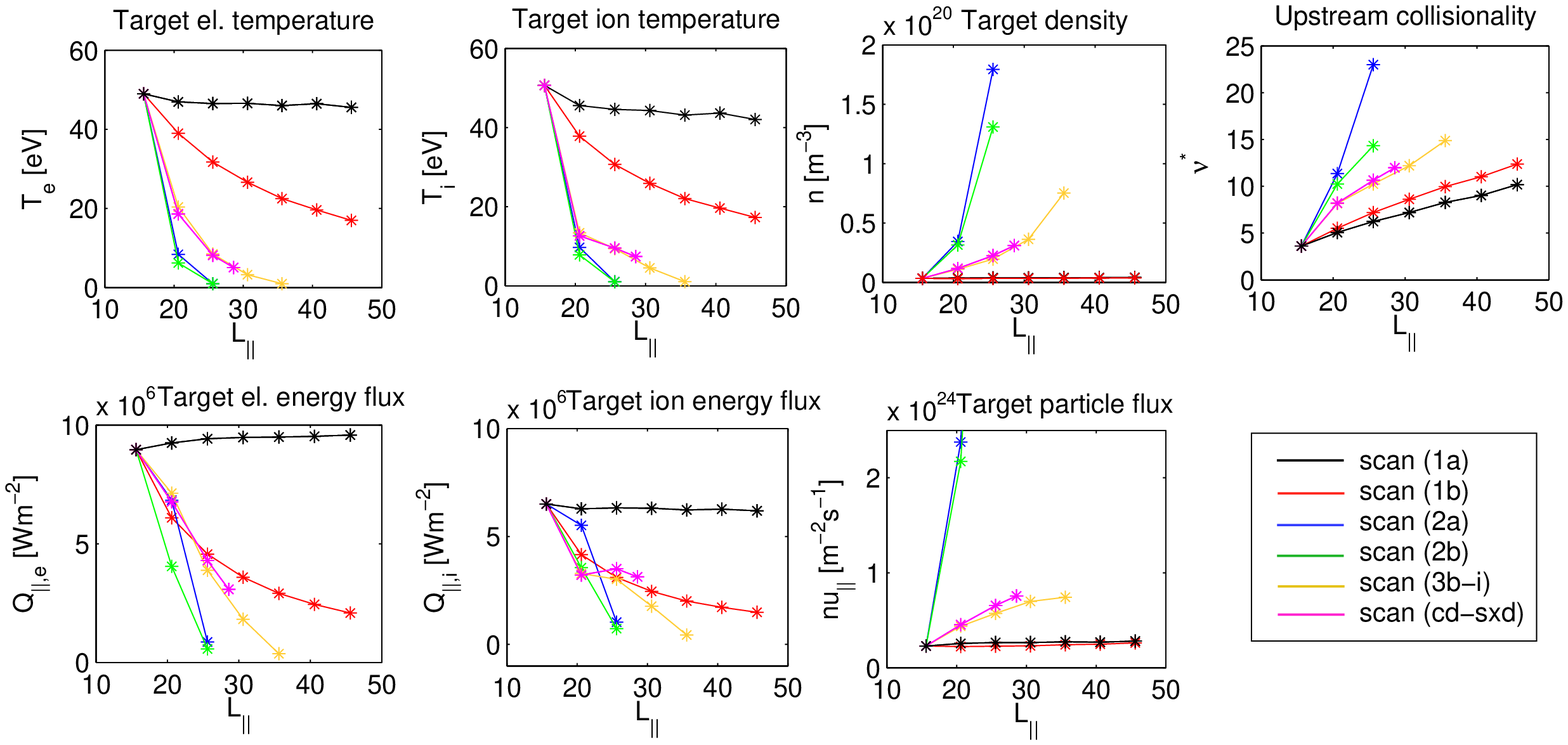}}}
\caption{Parameters at the target calculated by SOLF1D as functions of the connection length for lower temperatures. The scans involve stretching of the divertor region as described in the text. (1a) Constant integral sources in the flux tube (i.e. sources in the divertor are reduced proportionally to the increased divertor length), constant magnetic field in the divertor (see Fig. \ref{fig_stretch} a). (1b) Constant integral sources, varying magnetic field (Fig. \ref{fig_stretch} b). (2a) Sources in the divertor multiplied by the divertor length, constant magnetic field in the divertor. (2b) Sources in the divertor multiplied by the divertor length, varying magnetic field. (3b-i) Sources in the divertor varied as $S\propto L_{\rm div}^{\alpha}$, the magnetic field as in case (b). (cd-sxd) Sources in the divertor varied as $S\propto L_{\rm div}^{\alpha}$, the magnetic field as in SXD case.}
\label{fig_scan1d_2}
\end{figure}

As MAST-U SOLPS simulations can not yet be quantitatively compared with experiment, a benchmark simulation has been carried out for available MAST discharges. For similar radial heat conductivities as used in this paper, SOLPS tends to overestimate the plasma temperatures at the separatrix. While in the H-mode simulation $T_{\rm e,sep}\approx 100$ eV with no $T_{\rm e}$ drop towards the targets (sheath-limited), the experimental value is typically $T_{\rm e,sep}\approx 50$ eV and at the target $T_{\rm e,t}\approx 20$ eV. Because the volumetric power losses in the SOL are stronger for larger collisionality, the temperature (or input power to the flux tube) is a relevant parameter influencing our analysis. While the scan in Fig. \ref{fig_scan1d} remains useful for prediction of the target parameters for large $L_{\parallel}$ at high temperatures, we repeat the same analysis for temperatures that are more likely to occur in the early stages of MAST-U. The scan in Fig. \ref{fig_scan1d_2} is based on a SOLPS simulation for the CD case with lower input power to the SOL. Although $T_{\rm e}$ drops by a factor of 2, the SOL is still in the sheath-limited regime, with a slight gradient in $T_{\rm e}$ toward the target.

\begin{figure}[!h]
\centerline{\scalebox{0.63}{\includegraphics[clip]{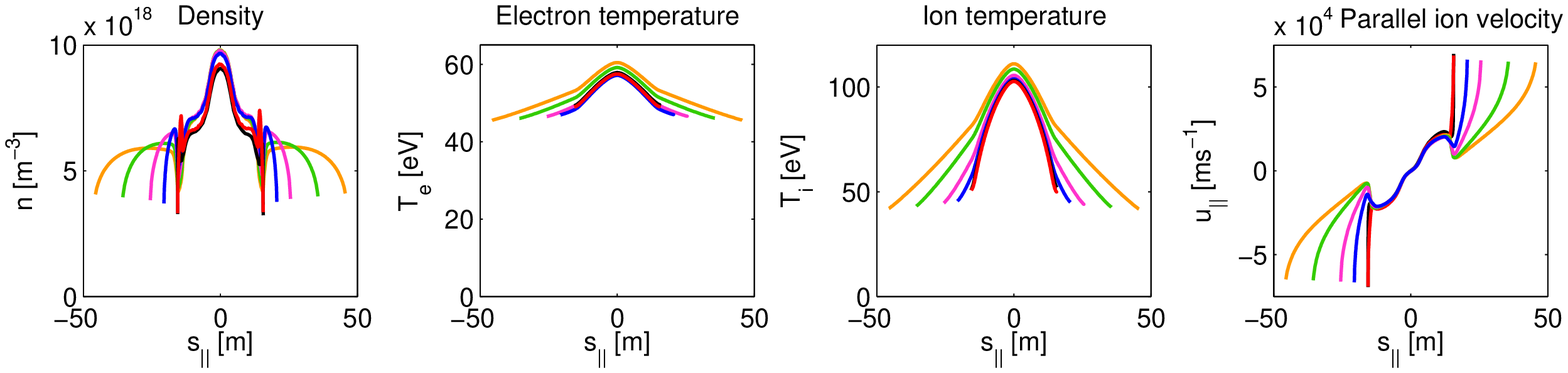}}}
\centerline{\scalebox{0.63}{\includegraphics[clip]{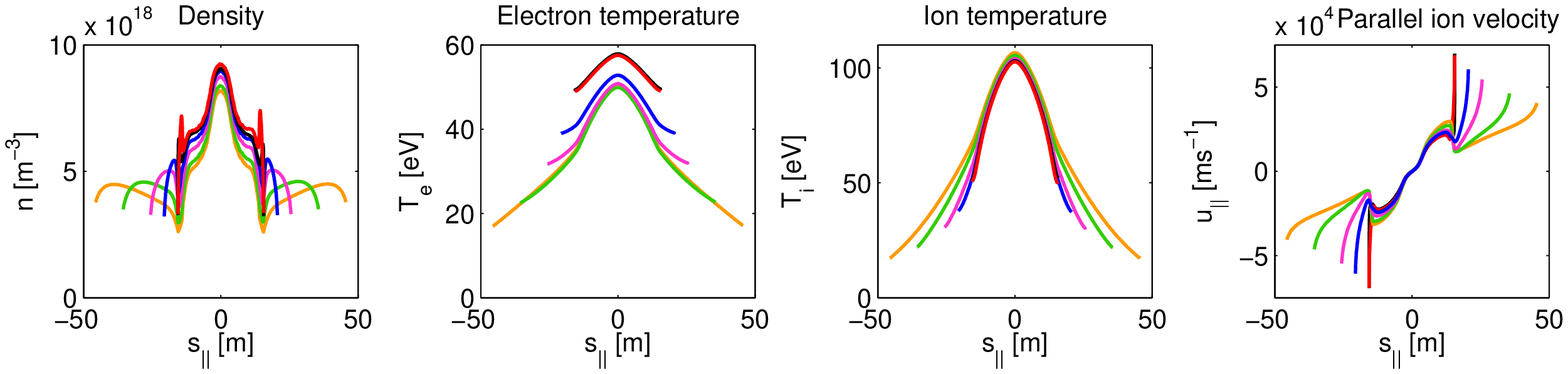}}}
\centerline{\scalebox{0.63}{\includegraphics[clip]{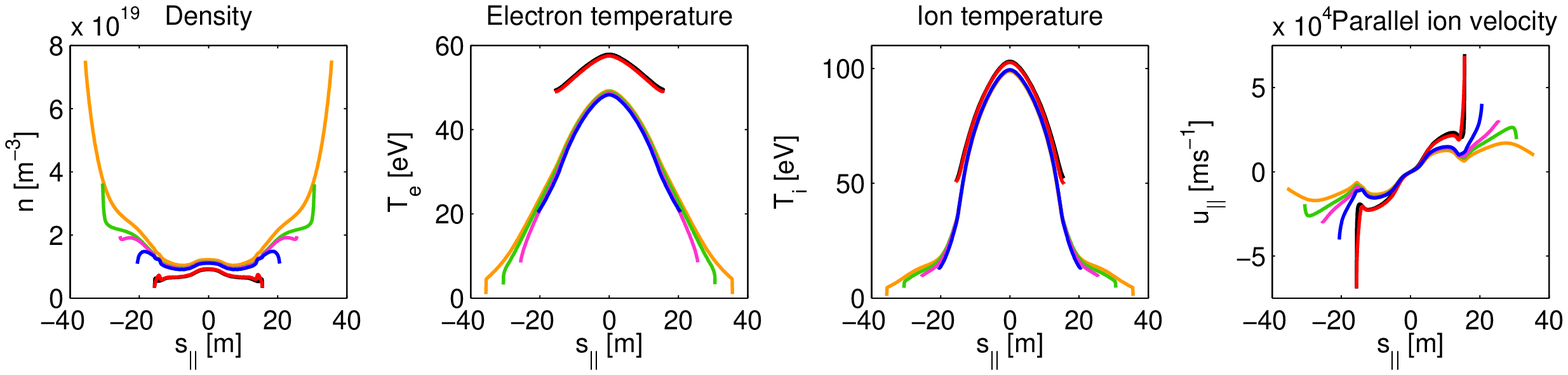}}}
\centerline{\scalebox{0.63}{\includegraphics[clip]{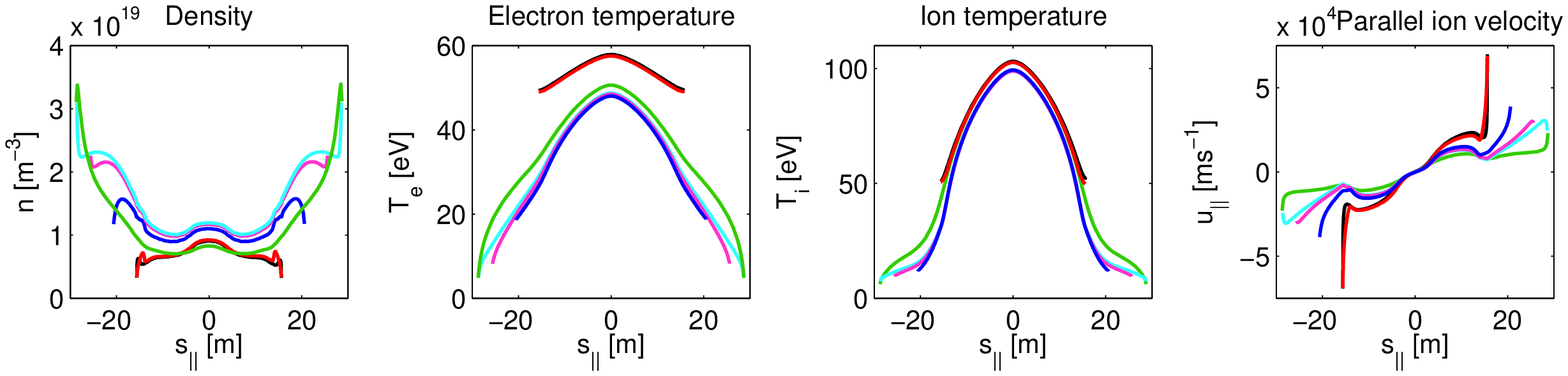}}}
\caption{Parallel profiles of plasma quantities on a flux tube between upper and lower outer targets for an expanded divertor leg at low temperatures: (i) Case (1a) from Fig. \ref{fig_scan1d_2}. (ii) Case (1b) from Fig. \ref{fig_scan1d_2} (increasing flux expansion). (iii) Case (3b-i) from Fig. \ref{fig_scan1d_2} (increasing flux expansion and the strength of recycling sources). In black, profiles for the CD geometry. In red, a solution for the CD geometry with the recycling sources at the target replaced uniformly by the average value in the divertor. In blue, magenta, green and orange, solutions for the expanded divertor. (iv) Case (cd-sxd) from Fig. \ref{fig_scan1d_2} (changing flux expansion and the strength of recycling sources as in SXD). In black, profiles for the CD geometry. In blue, magenta and cyan, profiles for the expanded divertor. The last profile with $L_{||}=28$ m coincides at the targets with the solution for the SXD geometry shown in green. 
}
\label{fig_profiles1d_2}
\end{figure}

At lower temperatures ($T_{\rm e}\approx 50$ eV at the target in CD), method (2b) results in the transition to detachment at $L_{\parallel}\approx 20-25$ m (the method with strong cooling in the divertor). Method (3b-i) leads to detached conditions ($T_{\rm e}$ below 5 eV) at $L_{\parallel}\approx 29$ m (the method that approximates best the transition from CD to SXD as simulated by SOLPS), i.e. at lower temperature, we need a shorter $L_{\parallel}$, by approximately 15 m, to detach. Also volumetric power losses are stronger, showing their impact on reducing the target loads below the drop observed from flux expansion alone at $L_{\parallel}\approx 25-30$ m.  

Full parallel profiles in the stretched divertor leg are shown in Fig. \ref{fig_profiles1d_2} which can be compared with the high-temperature scan in Fig. \ref{fig_profiles1d}. The first row is the reference case (1a) where no drop of target temperatures or fluxes is observed. The electron temperatures have only small gradients towards the target compared to ions with smaller parallel heat conductivity ($\kappa \propto m^{-1/2}$), but larger gradients compared to the previous case in Fig. \ref{fig_profiles1d} ($\kappa \propto T^{5/2}$). The second row (1b) shows the reduction of temperatures with flux expansion. Here, not only the target $T_{\rm e}$ drops with reduced energy flux, but also larger $T_{\rm e}$ gradients towards the target develop as the parallel electron conductivity drops. Additional ionization and collisional cooling in the divertor in the third row (3b-i) lead to a further drop of the target temperatures down to detachment at $L_{\parallel}\approx 30$ m. The bottom row shows how SOLF1D/SOLPS simulations of CD (black) and SXD (green) fit into the 1D scan, where recycling sources are for simplicity uniformly spread in the divertor. At $L_{\parallel}\approx 15$ m and $L_{\parallel}\approx 28$ m, the scan (red and cyan) recovers accurately the target values found in CD and SXD. 

\begin{figure}[!h]
\centerline{\scalebox{0.63}{\includegraphics[clip]{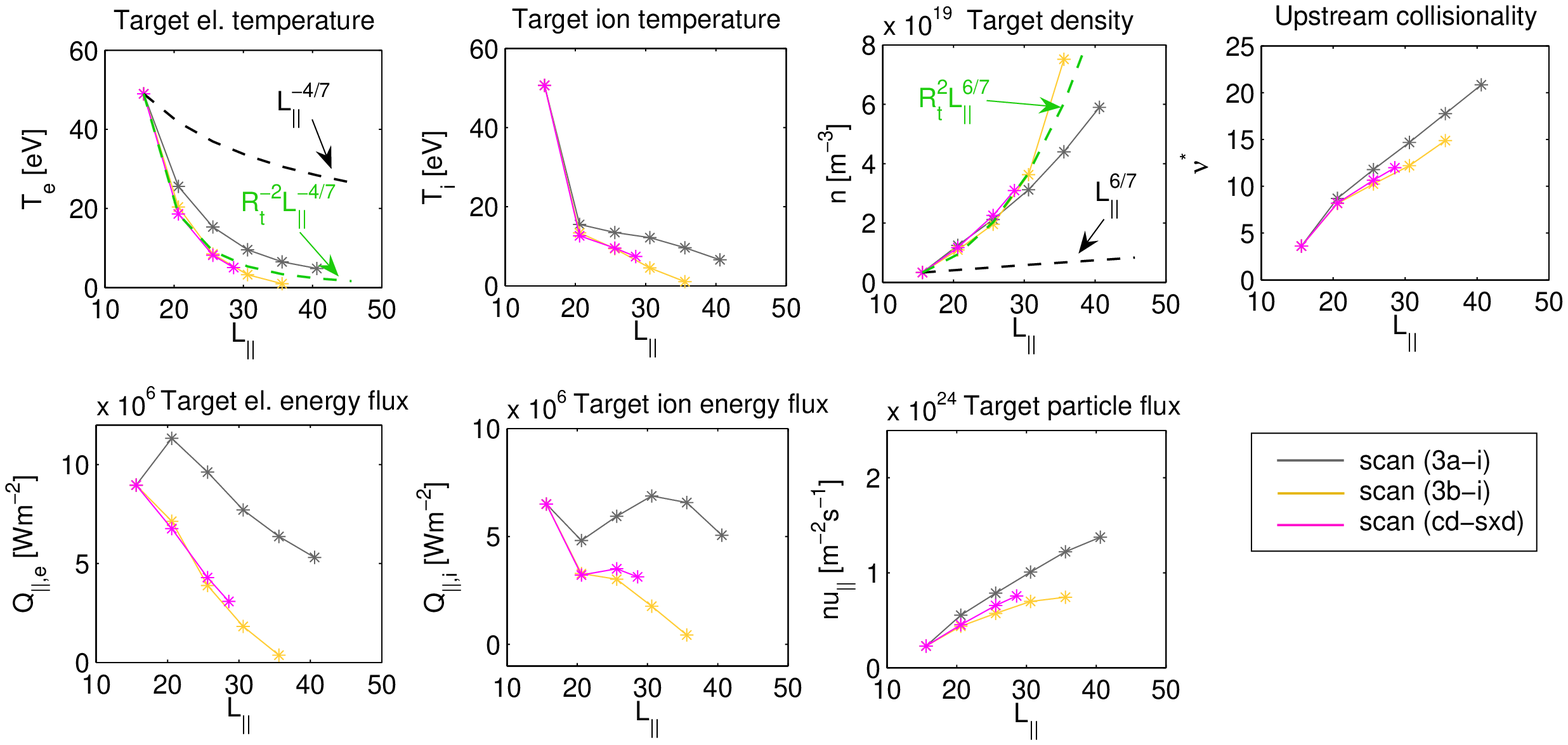}}}
\caption{Parameters at the target calculated by SOLF1D at lower temperatures compared to the two-point model. (3a-i) Constant magnetic field in the divertor, sources in the divertor vary as $S\propto L_{\rm div}^{\alpha}$ and match the sources in CD and SXD at $L_{||}\approx 15$ m and $L_{||}\approx 28$ m. (3b-i) As in Fig. \ref{fig_scan1d_2}, i.e. the magnetic field in the divertor drops as $B\propto 1/R$, sources in the divertor go as $S\propto L_{\rm div}^{\alpha}$. (cd-sxd) As in Fig. \ref{fig_scan1d_2}, the magnetic field and sources change as in SXD, therefore plasma parameters at $L_{||}\approx 15$ m coincide with the CD values and at $L_{||}\approx 28$ m with the SXD values. For comparison, the two-point model without/with $R_{\rm t}$ dependence is shown in black/green.}
\label{fig_2p_2}
\end{figure}

As in section \ref{2p}, the temperature drop and density increase are compared with the two-point model in Fig. \ref{fig_2p_2}. 
While the temperature drop is much stronger in the simulation for the case when the divertor is extended vertically (grey) than predicted by the two-point model (black), the decay to low temperatures in case of the radial stretching (yellow) is not far from the two-point model (green), although based on different physics. 

The last comment is related to -- (i) cooling due to sputtered impurity, (ii) drift effects -- two effects which are not included in the analysis, but which can be tested in SOLPS. If carbon sputtering is taken into account (with the chemical sputtering yield of 1\%), the expected effect on $L_{\parallel}$ at which the transition to the detachment occurs is not large. The temperature at the target is reduced by approximately 2 eV thanks to the impurity cooling for a case close to the detachment and one would gain approximately 2 m of the connection length, i.e. the detachment is expected at $L_{\parallel}\approx 27$ m with the impurity cooling in comparison to previous $L_{\parallel}\approx 29$ m without carbon.

Drift effects have not been tested in the frame of this paper, however, they have been studied for MAST separately in \cite{Vladimir1}. The key role is played by the poloidal ${\bf \rm E\times B}$ drift, which in a connected double null configuration results in an asymmetry of the plasma parameters in the top and bottom divertors. For the direction of the drift towards the lower outer plate, the temperatures at the lower outer plate are increased, while they are reduced at the upper outer target with respect to a symmetric situation without drifts. In \cite{Vladimir1}, the change in $T_{\rm e}$ varies between 0--50\% at the outer targets, while the largest effect is found in the inner lower divertor, where $T_{\rm e}$ can be reduced by up to a factor of 6. 
In our analysis in Fig. \ref{fig_scan1d_2}, the asymmetry resulting from the drift effects would cause that the predicted $L_{\parallel}$ at which the transition to the detachment occurs would be longer in the bottom divertor, while an earlier transition (at shorter $L_{\parallel}$) would be found at the top target. In \cite{Vladimir2}, the effect is quantified for SXD in MAST-U for low target temperatures of approximately 4 eV. One can see a drop of the peak $T_{\rm e}$ at the top from 4 to 1.8 eV and only a little rise of $T_{\rm e}$ at the bottom from 4 to 4.8 eV. With the change in the target temperature of few eV caused by the drifts, one should not expect the effect of the drifts on estimated $L_{\parallel}$ to be larger than 5 metres. 

\section{Conclusions}
In order to use the SOLF1D code to predict conditions at the target for a long-legged divertor, the code has first been benchmarked with SOLPS. The comparison shows a good agreement for the collisional SOL and a satisfactory agreement for the sheath-limited SOL, with discrepancies partly caused by numerical errors related to the discretization of SOLPS equations on a 2D grid and partly by an inaccurate conservation of the poloidal magnetic flux on the flux surface, once discretized on the grid. These discrepancies are, however, not critical (up to 20\%) and might be reduced by a higher resolution of the 2D grid. It has also been found that the SOLPS boundary condition for $T_{\rm i}$ at the divertor plate (omitting the kinetic part of the ion flux) leads to approximately 20\% lower $T_{\rm i}$ compared to SOLF1D. 

Based on the successful benchmark of SOLF1D with SOLPS for two divertor geometries (the conventional and Super-X divertor), an extrapolation to larger $L_{\parallel}$ is carried out. The effect of magnetic flux expansion on the reduction of the target temperature and energy flux is separated from the effect of power losses due to atomic processes. This is done by stretching the divertor leg in either radial ($B\propto 1/R$) or vertical (no additional flux expansion) directions. For a given initial value of $T_{\rm e,sep}\approx 112$ eV, $T_{\rm e}$ in front of the target drops from 110 eV to 25 eV 
between the conventional divertor geometry ($L_{\parallel}\approx 15$ m) and SXD ($L_{\parallel}\approx 28$ m), and for $L_{\parallel}\approx 45$ m (when the divertor leg is extended in the radial direction), $T_{\rm e}$ is further reduced to 6 eV. The temperature is reduced equally by the magnetic flux expansion and the plasma-neutral collision effects in this collisionality regime ($n_{\rm sep}\approx 9\times 10^{18}$ m$^{-3}$, $T_{\rm e,sep}\approx 112$ eV, $P_{\rm inp}\approx 1.7$ MW) and the $T_{\rm e}$ drop in SXD is twice slower than predicted by the two-point model. 

$Q_{\parallel}$ drops from 38.6 MWm$^{-2}$ (CD) to 25.1 MWm$^{-2}$ (SXD) under attached divertor conditions and for $L_{\parallel}\approx 45$ m, an additional drop to 11.7 MWm$^{-2}$ is predicted. The dominant effect responsible for the reduction of $Q_{\parallel}$ at the target is the magnetic flux expansion, while the volumetric power loss in the divertor starts to play a role at large $L_{\parallel}$ only ($L_{\parallel}\approx 45$ m), unless the radiation is increased at lower $L_{\parallel}$ by other means (e.g. increased density, impurities). The reason is the small importance of the collision-based power losses at small $L_{\parallel}$ in our case with respect to the energy source due to upstream cross-field transport. Moreover, the simulation shows that in a near-separatrix flux tube, the cross-field energy source increases from CD to SXD (due to larger parallel $\nabla T$ and stronger parallel heat transport), explaining why $Q_{\rm \parallel,e}$ does not drop in SXD (the stronger cross-filed source cancels the effect of stronger flux expansion and cooling). It is therefore desirable to increase the collisionality (i.e. radiation power losses) and reach detachment in order to achieve a further drop of $Q_{\parallel}$ via recombination and a reduction of $\Gamma_{\parallel}$ at the same time. 

Current MAST experiments show a smaller value of $T_{\rm e,sep}$ than the temperatures found in simulations, therefore we extrapolate the behaviour in target parameters at large $L_{\parallel}$ for a lower temperature as well. This adds one more parameter in our analysis (the collisionality) and appears to be relevant as the volumetric power losses in the SOL become more important. Compared to the previous scan at $T_{\rm e,sep}\approx 112$ eV, we gain approximately 15 m of the connection length. The target temperature drops from $T_{\rm e}\approx 49$ eV in the short divertor configuration ($n_{\rm sep}\approx 9\times 10^{18}$ m$^{-3}$, $T_{\rm e,sep}\approx 58$ eV, $P_{\rm inp}\approx 0.84$ MW), down to the detachment limit at $L_{\parallel}\approx 25-30$ m, hence SXD would be just around the detachment limit. A stronger reduction of $Q_{\rm \parallel,e}$ due to collisional cooling in this case is obvious.

Out of scope of the 1D analysis is the assessment of the poloidal flux expansion and the effect of the target tilting -- additional channels for reducing the energy flux deposited at the target, which can be expressed in terms of the parallel flux as $Q_{\rm t}=Q_{\parallel}(B_{\rm pol}/B)_{\rm t}{\rm sin}\beta$, $\beta$ is the tilting angle and $(B_{\rm pol}/B)_{\rm t}$ relates to the local pitch angle. 
From equilibrium and SOLPS calculations for the two divertor configurations of MAST-U, CD and SXD (where the poloidal magnetic field in the divertor is reduced by additional divertor coils), the poloidal flux expansion accounts for a factor of 2 decrease of the target energy load. 


\section*{Acknowledgments}
This work was supported by EURATOM and carried out within the framework of the European Fusion Development Agreement. The views and opinions expressed herein do not necessarily reflect those of the European Commission. The CCFE authors were funded by the RCUK Energy Programme under grant EP/I501045. 




\clearpage

\pagebreak
\section*{Appendix -- SOLF1D model}
\label{sec_app}


\subsection{Generalized equations} 
The model is based on the following equations for the plasma density $n$, parallel ion velocity $u_{\parallel}$ and electron and ion temperatures $T_{\rm e}$ and $T_{\rm i}$
\begin{eqnarray}
\hspace{-2cm}\frac{\partial n}{\partial t}+B\frac{\partial}{\partial s}\left(\frac{nu_{\parallel}}{B}\right)=S^n, \label{eq_continuity2} \\
\hspace{-2cm}\frac{\partial}{\partial t}(m_{\rm i}nu_{\parallel})+B\frac{\partial}{\partial s}\left(\frac{m_{\rm i} n u_{\parallel}^2}{B}\right)+B^{\frac{3}{2}}\frac{\partial}{\partial s}\left(B^{-\frac{3}{2}} \delta p_{\rm i}\right)=-\frac{\partial}{\partial s}(p_{\rm e}+p_{\rm i})+m_{\rm i}S^u, \label{eq_momentum2} \\
\hspace{-2cm}\frac{\partial}{\partial t}\left(\frac{3}{2}nkT_{\rm e}\right)+B\frac{\partial}{\partial s}\left[\frac{1}{B}\left(\frac{5}{2}nkT_{\rm e}u_{\parallel}+q_{\rm \parallel,e}\right)\right]=u_{\parallel}\frac{\partial p_{\rm e}}{\partial s}+Q_{\rm e}+S_{\rm e}^E \label{eq_energy12} \\
\hspace{-2cm}\frac{\partial}{\partial t}\left(\frac{3}{2}nkT_{\rm i}+\frac{1}{2}m_{\rm i}nu_{\parallel}^2\right)+B\frac{\partial}{\partial s}\left[\frac{1}{B}\left(\frac{5}{2}nkT_{\rm i}u_{\parallel}+\frac{1}{2}m_{\rm i}nu_{\parallel}^3+q_{\rm \parallel,i}+u_{\parallel}\delta p_{\rm i} \right)\right]= \label{eq_energy22} \\
-u_{\parallel}\frac{\partial p_{\rm e}}{\partial s}+Q_{\rm i}+S_{\rm i}^E  \nonumber
\end{eqnarray}
assuming quasineutrality ($n_{\rm e}=n_{\rm i}$), no net parallel current ($j_{\parallel}=0$) and ambipolarity ($u_{\rm \parallel,e}=u_{\rm \parallel,i}$). We further assume the generalized Ohm's law for electron momentum $enE_{\parallel}=-\partial p_{\rm e}/\partial s+R_{\rm \parallel,e}$ which leads to the cancellation of terms with thermal forces $R_{\rm \parallel,e}=-R_{\rm \parallel,i}$ and parallel electric field $E_{\parallel}$ in the momentum and energy equations and the substitution by $\partial p_{\rm e}/\partial s$ term, see Eqs. \ref{eq_continuity}-\ref{eq_energy2}.



\subsection{Relation to Braginskii model}
The set of equations is consistent with Braginskii equations, see \cite{Zawaideh1}. The divergence of the velocity vector ${\bf u}$ in the three-dimensional continuity equation of Braginskii (or analogically the energy flux vector in the energy equation) is replaced by
\begin{equation}
\nabla \cdot {\bf u} \rightarrow \frac{\partial u_{\parallel}}{\partial s} - \frac{u_{\parallel}}{B}\frac{\partial B}{\partial s} = B \frac{\partial}{\partial s}\left(\frac{u_{\parallel}}{B}\right), \label{eq_operator}
\end{equation}
leading to Eq. (\ref{eq_continuity2}). We obtain the new operator in (\ref{eq_operator}) by expanding the velocity as ${\bf u} = u_{\parallel}{\bf b} + {\bf u_{\perp}}$ and using $\nabla \cdot {\bf B}=0$ where ${\bf b} = {\bf B}/B$. The component perpendicular to the magnetic field will appear as a source term on the right-hand side of the one-dimensional equation. 


The momentum equation, Eq. (\ref{eq_momentum2}), is the parallel component of Braginskii momentum equation. Braginskii viscosity tensor ${\bf \pi} \equiv {\bf P}-p{\bf I}$ yields
\begin{equation}
{\bf \pi} = \delta p_{\rm i}\left(\frac{3}{2}{\bf b}{\bf b}-\frac{1}{2}{\bf I}\right), \label{eq_tensor}
\end{equation}
if following definitions are used ${\bf P} \equiv p_{\parallel}{\bf b}{\bf b}+p_{\perp}({\bf I}-{\bf b}{\bf b})$, $p \equiv (p_{\parallel}+2p_{\perp})/3$ and $\delta p_{\rm i} \equiv p_{\parallel}-p$. The parallel component of the viscous term in Braginskii momentum equation is then equivalent to
\begin{equation}
{\bf b} \cdot \nabla \cdot {\bf \pi} = \frac{\partial \delta p_{\rm i}}{\partial s}-\frac{3}{2}\frac{\delta p_{\rm i}}{B}\frac{\partial B}{\partial s} = B^{\frac{3}{2}}
\frac{\partial}{\partial s} \left(B^{-\frac{3}{2}}\delta p_{\rm i}\right).
\end{equation}

In order to complete the description of the model, a closure of the equations is required (i.e. expressions for the viscous momentum flux $\delta p_{\rm i}$ and the thermal heat fluxes $q_{\rm \parallel,e}$ and $q_{\rm \parallel,i}$) and all transport coefficients and source terms arising due to plasma-neutral interactions must be defined. 

\subsection{Parallel ion viscosity}
\label{sec_viscosity}

The parallel viscous flux is written as
\begin{equation}
\delta p_{\rm i} \approx -\eta_{\parallel} \left(\frac{\partial u_{\parallel}}{\partial s}+\frac{u_{\parallel}}{2B}\frac{\partial B}{\partial s}\right) = -\eta_{\parallel}B^{-\frac{1}{2}}\frac{\partial}{\partial s}\left(B^{\frac{1}{2}}u_{\parallel}\right). \label{eq_viscosity}
\end{equation}
Note that the form of Eq. (\ref{eq_viscosity}) is again consistent with Braginskii parallel viscous momentum flux, if we replace $\nabla \cdot {\bf u}$ by Eq. (\ref{eq_operator}) in individual components of Braginskii viscosity tensor. We can use the classical Braginskii parallel viscosity
\begin{equation}
\eta_{\parallel}=\eta_{\rm cl}=0.96nkT_{\rm i}\tau_{\rm i} \label{eq_eta}
\end{equation}
or employ viscous flux limiters in order to satisfy all collisionality regimes. The collision time $\tau_{\rm i}$ is defined as
\begin{equation}
\tau_{\rm i}=\frac{3\sqrt{m_{\rm i}}(kT_{\rm i})^{\frac{3}{2}}}{4\sqrt{\pi}n\lambda e^4}=2.09 \times 10^{13} \frac{T_{\rm i}^{\frac{3}{2}}}
{n\lambda}\sqrt{\frac{m_{\rm i}}{m_{\rm p}}} {\rm sec}
\end{equation}
in SI units and Boltzmann constant is $k=1.6 \times 10^{-19}$ J/eV \cite{NRL}. The Coulomb logarithm $\lambda$ is generally a function of the density and temperature, see e.g. \cite{Wesson} or \cite{Stangeby}.

If we follow the derivation of the parallel and perpendicular pressure equations as done for example in \cite{Zawaideh1} or \cite{Wojtek1}, we can obtain more general equation for the parallel viscous flux $\delta p_{\rm i}$ which defines a parallel viscosity limiter through
\begin{equation}
\eta_{\parallel}=\frac{\eta_{\rm cl}}{1+\Omega_{\eta}}, \quad
\Omega_{\eta}=\frac{\eta_{\rm cl}\nabla_{\parallel}u_{\parallel}}{\frac{4}{7}nkT_{\rm i}}-\frac{\eta_{\rm cl} u_{\parallel}\nabla_{\parallel}B}{BnkT_{\rm i}}. \label{eq_visc}
\end{equation}
Eq. (\ref{eq_visc}) reduces to the expression (\ref{eq_eta}) in the limit of high collisionality. In widely used 2D transport codes such as SOLPS or EDGE2D, only the $\nabla_{\parallel}u_{\parallel}$ term is taken into account, with the ion viscous flux limiter as an optional parameters in the code, but typically $\beta_{u}=0.5$ ($\approx 4/7$) being a good choice in steady-state inter-ELM modelling of the low-recycling SOL. 

\subsection{Parallel heat conductivity}
From higher-order moment equations, approximate expressions for the heat flux can be obtained. We calculate the heat flux using classical Spitzer-H\"arm heat conductivities 
\begin{eqnarray}
q_{\rm \parallel,e}=-\kappa_e\frac{\partial}{\partial s}(kT_{\rm e}), \quad \kappa_{\rm e}=\kappa_{\rm cl,e}=3.2\frac{nkT_{\rm e} \tau_{\rm e}}{m_{\rm e}}, \\
q_{\rm \parallel,i}=-\kappa_i\frac{\partial}{\partial s}(kT_{\rm i}), \quad \kappa_{\rm i}=\kappa_{\rm cl,i}=3.9\frac{nkT_{\rm i} \tau_{\rm i}}{m_{\rm i}},
\end{eqnarray}
and define the electron collision time as
\begin{equation}
\tau_{\rm e}=\frac{3\sqrt{m_{\rm e}}(kT_{\rm e})^{\frac{3}{2}}}{4\sqrt{2\pi}n\lambda e^4}=3.44 \times 10^{11} \frac{T_{\rm e}^{\frac{3}{2}}}
{n\lambda} {\rm sec}.
\end{equation}

Kinetic corrections in the form of heat flux limiters can be used. SOLF1D allows to modify the thermal heat flux both for the electrons and ions as
\begin{equation}
q_{\parallel} =\left(\frac{1}{q_{\rm \parallel,lim}}+\frac{1}{q_{\rm \parallel,cl}}\right)^{-1} \label{eq_hflim}
\end{equation}
which limits the heat flux to a maximum acceptable value $q_{\rm \parallel,lim}=\alpha n v^{\rm th} kT$, where $v^{\rm th}$ is the thermal speed, and imposes a limit for the heat conductivity which would otherwise diverge for large temperatures. From Eq. (\ref{eq_hflim}), a corrected expression for the heat conductivity can be formulated as
\begin{equation}
\kappa = \frac{\kappa_{\rm cl}}{1+\Omega_{\kappa}}, \quad
\Omega_{\kappa}=\frac{\kappa_{\rm cl}\nabla_{\parallel}T}{\alpha nv^{\rm th}T}.
\end{equation}
The electron and ion heat flux limiters $\alpha_{\rm e}$ and $\alpha_{\rm i}$ are again optional parameters of the model. As a result of kinetic studies, values of the heat flux limiters are observed in the range $0.03 \leq \alpha \leq 0.6$ with poloidally averaged values $\alpha \approx 0.15 \pm 0.05$ (depending on the collisionality) and for the viscosity limiter it is $\beta \approx 0.5 \pm 0.1$ \cite{David1}. At high collisionalities, no limiting is required and some authors mention a heat flux enhancement contrary to limiting \cite{Alex1}. The limiters strongly vary in time, e.g. during ELMs or turbulent transport, by several orders of magnitude \cite{David1} and the latest comparison of SOLF1D with the kinetic code BIT1 has shown that assuming constant heat flux limiters during the ELM crash is not adequate \cite{Eva5}.

\subsection{Model of neutrals}

Atomic species are treated as a fluid and their transport is described by 1D continuity and momentum transfer equations
\begin{eqnarray}
\frac{\partial n_{\rm 0}}{\partial t}+\frac{\partial}{\partial s}\left(n_{\rm 0} u_{\rm 0}\right)=S_{\rm 0}^n, \\ 
\frac{\partial}{\partial t}\left(m_{\rm 0}n_{\rm 0}u_{\rm 0}\right)+\frac{\partial}{\partial s}\left(m_{\rm 0}n_{\rm 0}u_{\rm 0}^2\right)=-\frac{\partial p_{\rm 0}}{\partial s}+m_{\rm 0}S_{\rm 0}^u
\end{eqnarray}
with the density and momentum sources/sinks $S^n_0$ balanced by corresponding ionic sinks/sources in the plasma fluid equations. 
The closure is made using an assumption about the energy of neutrals. Neutral species are assumed to thermally equilibrate with ions due to dominant charge-exchange processes and therefore considered to have the temperature locally equal to the ion temperature $T_0=T_{\rm i}$ everywhere in the SOL.

The 1D model of neutrals provides a simple way to incorporate the main aspects of the SOL for different collisionalities and to describe high-recycling conditions. While the 1D description is reasonable for plasma, 2D modelling of neutrals would be more appropriate, especially if the ionization mean free path is long enough for neutrals to propagate deeper in the SOL. In such case, 1D model can lead to overestimation of neutral concentration on the flux tube or eventually result in instable solutions. 2D codes such as SOLPS are usually coupled with Monte Carlo EIRENE and there has been an evidence that the kinetic treatment is certainly required for precise quantitative calculations (see e.g. \cite{DC1}). Both plasma and neutral models in SOLF1D are currently being benchmarked with PIC simulations performed with BIT1 code for different collisionalities and results will be published shortly.

\subsection{Collision and source terms}

\begin{figure}[!h]
\centerline{\scalebox{0.6}{\includegraphics[clip]{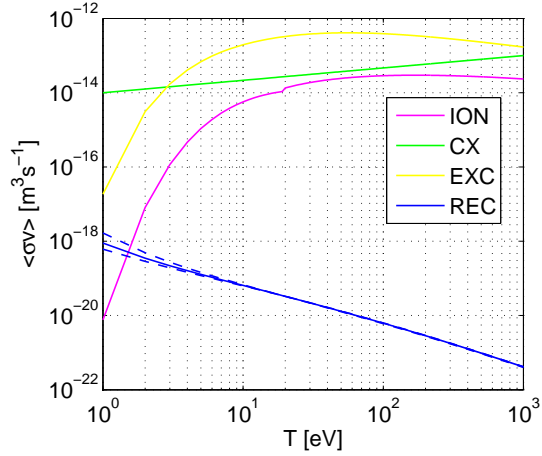}}}
\caption{Collision rates used in the SOLF1D code. The recombination rate is displayed for densities $n=1\times 10^{18}$ m$^{-3}$, $n=1\times 10^{19}$ m$^{-3}$ and $n=1\times 10^{20}$ m$^{-3}$.} \label{fig_atomdata}
\end{figure}

The energy exchange between electron and ions due to collisions is described as
\begin{equation}
Q_{\rm i} = - Q_{\rm e} = \frac{3m_{\rm e}}{m_{\rm i}}\frac{nk}{\tau_{\rm e}}(T_{\rm e}-T_{\rm i}).
\end{equation}
The sources $S^n$, $S^u$, $S_{\rm e}^E$ and $S_{\rm i}^E$ in Eqs. (\ref{eq_continuity2})-(\ref{eq_energy22}) comprise cross-field sources of plasma, momentum (here neglected) and energy and collision terms (the interaction of plasma with neutrals) describing changes of the mass, momentum and energy due to processes of ionization, charge exchange, excitation and recombination. They are defined as
\begin{eqnarray}
S^n=n_0n\langle \sigma v\rangle_{\rm ION}-n^2\langle \sigma v\rangle_{\rm REC}+S^n_{\perp}, \\
S^u=n_0nu_0\langle \sigma v\rangle_{\rm ION}+n_0n(u_0-u_{\parallel})\langle \sigma v\rangle_{\rm CX}-n^2u_{\parallel}\langle \sigma v\rangle_{\rm REC}, \\
S_{\rm e}^E=-n_0n\langle \sigma v\rangle_{\rm ION}kI_{\rm H}-n_0nkQ_{\rm H}+S^E_{\perp,e}, \\
S_{\rm i}^E=n_0n\langle \sigma v\rangle_{\rm ION}\left[\frac{3}{2}kT_0+\frac{1}{2}m_0u_0^2\right] \\
+n_0n\langle \sigma v\rangle_{\rm CX}\left[\frac{3}{2}k(T_0-T_{\rm i})+\frac{1}{2}m_0(u_0^2-u_{\parallel}^2)\right] \nonumber \\
-n^2\langle \sigma v\rangle_{\rm REC}\left[\frac{3}{2}kT_{\rm i}+\frac{1}{2}m_0u_{\parallel}^2\right]+S^E_{\perp,i}. \nonumber
\end{eqnarray}
$I_{\rm H}$ is the ionization potential ($I_{\rm H}=13.6$ eV for hydrogen ions), $Q_{\rm H}$ is the cooling rate due to excitation and $\langle \sigma v\rangle_{\rm ION}$, $\langle \sigma v\rangle_{\rm CX}$ and $\langle \sigma v\rangle_{\rm REC}$ are collision rates for ionization, charge exchange and recombination which are, in general, functions of the density and temperature, see Fig. \ref{fig_atomdata}. 

\subsection{Numerical solution}
The equations in SOLF1D are solved numerically in variables $n$, $u_{\parallel}$, $T_{\rm e}$ and $T_{\rm i}$, i.e. equations (\ref{eq_continuity2})--(\ref{eq_energy22}) are modified as

\begin{eqnarray}
\hspace{-2cm}\frac{\partial n}{\partial t}+B\frac{\partial}{\partial s}\left(B^{-1}nu_{\parallel}\right)=S^n, \\
\hspace{-2cm}m_{\rm i}n\frac{\partial u_{\parallel}}{\partial t}+\frac{1}{2}m_{\rm i}n\frac{\partial}{\partial s}(u_{\parallel}^2)-B^{\frac{3}{2}}\frac{\partial}{\partial s}\left[B^{-2}\eta_{\parallel}\frac{\partial}{\partial s}\left(B^{\frac{1}{2}}u_{\parallel}\right)\right]=\\
-\frac{\partial}{\partial s}(p_{\rm e}+p_{\rm i})+m_{\rm i}S^u-m_{\rm i}u_{\parallel}S^n, \nonumber \\
\hspace{-2cm}\frac{3}{2}nk\frac{\partial T_{\rm e}}{\partial t}+\frac{3}{2}nk\frac{\partial}{\partial s}(T_{\rm e}u_{\parallel})-B\frac{\partial}{\partial s}\left[B^{-1}\kappa_{\rm e}\frac{\partial}{\partial s}(kT_{\rm e})\right]=\\ \frac{1}{2}nkT_{\rm e}\frac{\partial u_{\parallel}}{\partial s}-\frac{3}{2}kT_{\rm e}S^n+Q_{\rm e}+S_{\rm e}^E+\frac{nkT_{\rm e}u_{\parallel}}{B}\frac{\partial B}{\partial s}, \nonumber \\ 
\hspace{-2cm}\frac{3}{2}nk\frac{\partial T_{\rm i}}{\partial t}+\frac{3}{2}nk\frac{\partial}{\partial s}(T_{\rm i}u_{\parallel})-B\frac{\partial}{\partial s}\left[B^{-1}\kappa_{\rm i}\frac{\partial}{\partial s}(kT_{\rm i})\right]=\\ \frac{1}{2}nkT_{\rm i}\frac{\partial u_{\parallel}}{\partial s}+\eta_{\parallel}\left(\frac{\partial u_{\parallel}}{\partial s}\right)^2-\frac{3}{2}kT_{\rm i}S^n+\frac{1}{2}m_{\rm i}u_{\parallel}^{2}S^n-m_{\rm i}u_{\parallel}S^u+Q_{\rm i}+S_{\rm i}^E \nonumber\\
+\frac{nkT_{\rm i}u_{\parallel}}{B}\frac{\partial B}{\partial s}+\frac{\eta_{\parallel}u_{\parallel}
}{B}\frac{\partial B}{\partial s}\frac{\partial u_{\parallel}}{\partial s}+\frac{\eta_{\parallel}u_{\parallel}^2}{4B^2}\left(\frac{\partial B}{\partial s}\right)^2, \nonumber
\end{eqnarray}
and solved in a similar form 
\begin{equation}
a\frac{\partial f}{\partial t}+b\frac{\partial}{\partial s}(fv)+c\frac{\partial}{\partial s}\left[d\frac{\partial}{\partial s}(ef)\right]=S_1+fS_2.
\end{equation}

The SOLF1D code was written as a simple and fast alternative to existing 2D codes, considering the attention to the numerical aspects less important. That is why the model is solved using traditional numerical methods of second order both in time and space. The system of nonlinear equations is solved by an algorithm based on the finite difference method. The equations are discretized on a non-uniform staggered grid using traditional numerical schemes and solved by a mixed explicit/implicit time integration. We use an exponential grid with refined spacing in the boundary regions where large gradients of plasma quantities can occur in high-recycling or detached regimes. 
The convective terms of the fluid equations are converted to finite difference expressions by the second-order upwind scheme and the diffusive terms are discretized by the Crank-Nicholson scheme. The time stepping is based on the second-order splitting method (see e.g. \cite{Karniadakis}). Nonlinear terms are treated explicitly, while linear terms are updated to a new time level implicitly. Resulting systems of linear equations are solved by the Progonka and Matrix Progonka methods described in \cite{Samarskii,Eva2}.

\subsection{Boundary conditions}

At both ends of the computational region (target plates), boundary conditions are applied (see Tab. \ref{tab_bc}), including Bohm criterion for the parallel ion velocity (Dirichlet boundary condition) and standard expressions for the sheath energy fluxes using constant sheath energy transmission factors (the condition for the flux is linearized and translated into Newton boundary condition for the temperature). A pumping at the target and neutral recycling is included using the recycling coefficient and recycled neutrals are assumed to propagate from the targets with the thermal speed at the temperature $T_0=T_{\rm i}$ (neutrals leaving the wall are assumed to equilibrate fast with plasma ions due to charge exchange). The density is extrapolated from the neighbouring points to the boundary. 

\begin{table}[h]
\begin{center}
\begin{tabular}{ll}
\hline \bfseries quantity &
\bfseries boundary condition \\
\hline $n$ & extrapolation \\
$u_{\parallel}$ & $u_{\parallel}=c_{\rm s}\equiv\sqrt{\frac{k(T_{\rm e}+T_{\rm i})}{m_{\rm i}}}$  \\
$T_{\rm e}$ & $Q_{\rm \parallel,e}=\delta_{\rm e}nkT_{\rm e}c_{\rm s}$  \\
$T_{\rm i}$ & $Q_{\rm \parallel,i}=\delta_{\rm i}nkT_{\rm i}c_{\rm s}$  \\
$n_{\rm 0}$ & $\Gamma_0=-R\Gamma_{\parallel}$  \\
$u_{\rm 0}$ & $u_0=v^{\rm th}\equiv\sqrt{\frac{kT_0}{m_0}}$ \\
\hline
\end{tabular}
\caption{Boundary conditions of the SOLF1D model.} \label{tab_bc}
\end{center}
\end{table}

\end{document}